\documentclass[aps,pra,epsf,superscriptaddress,amsmath,amssymb,amsfonts,twocolumn,showpacs]{revtex4-1}

\usepackage{graphicx}
\usepackage{epsfig}
\usepackage{dcolumn}
\usepackage{bm}
\usepackage{braket}
\usepackage{amsmath}
\usepackage{mathtools}
\usepackage{graphicx,color,xcolor}
\usepackage{hyperref}
\newcommand{\abs}[1]{\left| #1 \right|} 
\usepackage{hyperref}
\usepackage{multirow}
\usepackage[normalem]{ulem} %% REMOVE ME!

\begin{document}
\title{Radiofrequency spectroscopy of one-dimensional trapped Bose polarons:\\ 
crossover from the adiabatic to the diabatic regime}

\author{S. I. Mistakidis}
\affiliation{Center for Optical Quantum Technologies, Department of Physics, University of Hamburg, 
Luruper Chaussee 149, 22761 Hamburg Germany}
\affiliation{ITAMP, Center for Astrophysics $\vert$ Harvard $\&$ Smithsonian, Cambridge, MA 02138 USA} 
\author{G.M. Koutentakis}
\affiliation{Center for Optical Quantum Technologies, Department of Physics, University of Hamburg, 
Luruper Chaussee 149, 22761 Hamburg Germany} \affiliation{The Hamburg Centre for Ultrafast Imaging,
University of Hamburg, Luruper Chaussee 149, 22761 Hamburg, Germany}
\author{F. Grusdt}
\affiliation{Department of Physics and Arnold Sommerfeld Center for Theoretical Physics (ASC), 
Ludwig-Maximilians-Universitat M{\"u}nchen, Theresienstr. 37, M{\"u}nchen D-80333, Germany} 
\affiliation{Munich Center for Quantum Science and Technology (MCQST), Schellingstr. 4, D-80799 M{\"u}nchen, Germany}
\author{H. R. Sadeghpour}
\affiliation{ITAMP, Center for Astrophysics $\vert$ Harvard $\&$ Smithsonian, Cambridge, MA 02138 USA} 
\author{P. Schmelcher}
\affiliation{Center for Optical Quantum Technologies, Department of Physics, University of Hamburg, 
Luruper Chaussee 149, 22761 Hamburg Germany} \affiliation{The Hamburg Centre for Ultrafast Imaging,
University of Hamburg, Luruper Chaussee 149, 22761 Hamburg, Germany}

\date{\today}

\begin{abstract} 

We investigate the crossover of the impurity-induced dynamics, in trapped one-dimensional Bose polarons
subject to radio frequency (rf) pulses of varying intensity, from an adiabatic to a diabatic regime. 
Utilizing adiabatic pulses for either weak repulsive or attractive impurity-medium interactions, a multitude of 
polaronic excitations or mode-couplings of the impurity-bath interaction with the 
collective breathing motion of the bosonic medium are spectrally resolved. 
We find that for strongly repulsive impurity-bath 
interactions, a temporal orthogonality catastrophe manifests in resonances in the excitation spectra where impurity coherence vanishes. 
When two impurities are introduced, impurity-impurity correlations, for either attractive or strong repulsive couplings, induce a spectral shift of the resonances with respect to the single impurity. 
For a heavy impurity, the polaronic peak is accompanied by a series of equidistant side-band resonances, related to interference of the impurity spin dynamics and the sound waves of the bath.
In all cases, we enter the diabatic transfer regime for an increasing bare Rabi frequency of the 
rf field with a Lorentzian spectral shape featuring a single polaronic resonance. 
The findings in this work on the effects of external trap, rf pulse and impurity-impurity interaction should have implications for the new generations of cold-atom experiments. 

\end{abstract}

%\pacs{Pacs numbers} 
\maketitle

\section{Introduction} 

The concept of quasiparticles~\cite{Landau,Pekar} such as polarons~\cite{Massignan,Schmidt_rev}, 
mobile impurities dressed by their 
many-body (MB) host has found considerable interest in the cold-atom community, due to the precise tunability of inter-atomic interactions with tools such as Fano-Feshbach resonances~\cite{Chin,Kohler}. 
Spectroscopic techniques, such as rf and Ramsey 
spectroscopy~\cite{Koschorreck,Kohstall,Cetina,Cetina_interferometry,Camargo_rf} permit the characterization of quasiparticle states. 
Utilizing these probes and depending on the quantum statistics of the host both Bose~\cite{Jorgensen,Hu,Catani1,Fukuhara,Yan_bose_polarons} 
and Fermi~\cite{Scazza,Koschorreck,Kohstall} 
polarons have been recently realized mainly in multidimensional settings.   
The one-dimensional (1D) Bose and Fermi polarons are less explored~\cite{Meinert,Catani_1D,Wenz_pol}.

The above experimental findings have been corroborated by theory~\cite{Grusdt_approaches,Rath_approaches}, 
e.g. using variational approaches~\cite{Mistakidis_eff_mass,Mistakidis_orth_cat,Mistakidis_Fermi_pol,Ardila_MC,Ardila_res_int} and  
renormalization group methods~\cite{Grusdt_RG,Grusdt_strong_coupl}. 
These methods initially aimed to address the equilibrium polaron properties, such as their 
effective mass~\cite{Grusdt_1D,Ardila_MC,Jager_mass}, induced interactions~\cite{induced_int_artem,Mistakidis_Fermi_pol}, bound bipolaron~\cite{Bipolaron,Massignan,Schmidt_rev} and trimeron~\cite{Nishida_trim} states. 
More recently, the nonequilibrium dynamics of polarons~
\cite{Mistakidis_orth_cat,Mistakidis_two_imp_ferm,Mistakidis_eff_mass,Grusdt_RG,
Shchadilova,Kamar,Boyanovsky,Skou}
have attempted to address issues, such as the collisional properties with the 
host bath particles~\cite{Burovski_col,Lychkovskiy_col1,Meinert,Flutter,Flutter1,Gamayun_col,Mistakidis_inject_imp}, 
tunneling in optical lattices~\cite{Cai_transp,Johnson_transp,Siegl,Theel}, 
dynamics of doped insulators~\cite{Keiler_dopped,Bohrdt,Ji}, induced correlations~\cite{Mistakidis_induced_cor,Mistakidis_Volosniev_induce_int}, 
relaxation processes~\cite{Lausch_col1,Boyanovsky} and dynamical decay, i.e. the temporal orthogonality catastrophe (TOC)~\cite{Mistakidis_orth_cat,Mistakidis_induced_cor,Mistakidis_pump}. 

Here, we probe the dynamics of Bose polarons in 1D, where correlations can be enhanced, with the rf spectroscopy. The rf spectroscopy provides a powerful and well-established experimental toolkit~\cite{Jorgensen,Hu,Cetina,Gupta_rf} 
for monitoring the polaronic states and  the correlation dynamics. 
Recall that within a typical rf spectroscopy protocol, the MB system is prepared in its equilibrium state and subsequently a specific pulse drives the impurity atoms from their non-interacting to the interacting with the host spin state (injection spectroscopy) or vice versa (ejection spectroscopy)~\cite{Liu_rf1,Liu_rf2,Dzsotjan}. 

The rf scheme thus enables us to probe the polaron excitation at different response regimes of the impurities spin-flip dynamics, compared to the characteristic timescale of their atomic motion. 
It should be emphasized that irrespectively of the spatial dimension the crossover of the dynamical 
response of Bose polarons from  weak to strong Rabi coupling of the involved spin states has not been systematically studied. 
Therefore, it would be particularly interesting to unravel the dependence of the polaron states 
and their correlation properties with respect to the intensity of the rf pulse, a study that can 
potentially lead to controlling the polaron states. 
For instance, in the extreme limit of an intense rf field, the spin-transfer is much faster (diabatic) than the characteristic timescale of the 
atomic motion related to phononic emission.  
Then, we would expect that the spatial and spin degrees-of-freedom are decoupled 
and thus correlation is negligible. 
In contrast, weaker pulses lead to a slower (adiabatic) transfer when compared to the timescale set by the external confinement. 
As such, it is possible to retrieve information regarding 
the energy content of the system in terms of its eigenstates.  

A prototype system consists of one or two spinor impurity bosons embedded in a 1D harmonically trapped Bose-Einstein condensate (BEC). 
To monitor the excitation dynamics of this setting when subjected to an injection rf pulse, we deploy the variational nonperturbative method 
Multi-Layer Multi-Configuration Time-Dependent Hartree 
method for atomic mixtures (ML-MCTDHX)~\cite{MLX,MLB1,MLB2}, which enables us to capture the interparticle correlations. 

The formation of attractive or repulsive Bose polarons is considered for weakly attractive or repulsive impurity-medium coupling. 
More precisely, for pulses with a bare Rabi frequency smaller or equal to the phononic excitation frequency, a variety of polaronic resonances appear, reflecting the mode-coupling of the 
impurity motion to the collective breathing of the bosonic bath~\cite{Dolgirev}. 
The motionally excited polaronic states exhibit a decreasing quasi-particle residue. 
Remarkably, for strongly repulsive impurity-bath interactions, TOC of the Bose polaron~\cite{Mistakidis_orth_cat,Mistakidis_induced_cor,Mistakidis_pump}, 
 manifests in the spectrum which saturates for long pulse times. 
Here, the spectral response experiences a multitude of resonances corresponding to distinct superpositions of lower-lying energy states of the impurity. 
The rate of the polaron decay via TOC is found to be inherently faster for increasing bath-impurity spatial 
overlap, a process that can be controlled with detuning of the rf field. 
The suppressed coherence of 
the impurity in the course of the TOC means that the polaron also vanishes. 

In the case of a heavy impurity, 
irrespective of the impurity-medium interaction and the bare Rabi frequency (except the diabatic case), a polaronic resonance is always present, accompanied by a series of equidistant side peaks and troughs. 
This latter structure originates from the interference of the sound waves of the background with the impurity spin dynamics. 
Moreover regardless of the impurity-medium coupling and mass ratio, increasing the bare Rabi frequency (larger than the characteristic trap frequency), allows us to systematically probe the diabatic spin-transfer regime with the 
spectrum featuring a Lorentzian shape with a single polaronic resonance. 

For two non-interacting bosonic impurities and for either attractive or strong repulsive impurity-medium couplings, a small shift of the spectral resonance, compared with the single impurity case is observed. 
This fact elucidates the presence of induced impurity-impurity correlations mediated by the BEC medium~\cite{Mistakidis_induced_cor,Mistakidis_Volosniev_induce_int,Mistakidis_pump}.
In all cases, the resonance position can be tuned by varying the impurity-bath coupling or the bare Rabi frequency.

This work is organized as follows. 
Section~\ref{sec:theory} introduces the setup under consideration and the employed rf protocol to induce the 
impurity nonequilibrium dynamics. 
Additionally, the different theoretical approaches used for the interpretation of the impurity rf spectrum are outlined. 
In Sec.~\ref{sec:mass_balance} we analyze in detail the resultant spectral response for different impurity-medium interactions 
and importantly address the crossover from adiabatic to diabatic spin dynamics for different intensities of the rf pulse.  
The two-body processes contributing to the spectrum of two non-interacting impurities are discussed in Sec.~\ref{sec:two_body}. 
The emergent spectral response of an impurity heavier than the atoms of the corresponding medium for different impurity-BEC coupling and rf intensities is 
presented in Sec.~\ref{sec:heavy_imp}. 
We summarize and provide perspectives for future endeavors in Section~\ref{sec:conclusions}. 
Appendix~\ref{app:diabatic_approx} elaborates on the details of the diabatic approximation.

\section{Spectroscopy scheme and methological approaches}\label{sec:theory}

\subsection{Hamiltonian and radiofrequency protocol}\label{sec:hamiltonian_protocol}

To emulate the rf injection spectroscopy scheme~\cite{Jorgensen,Hu,Cetina,Shchadilova,Mistakidis_Fermi_pol}, we consider 
a highly particle imbalanced multicomponent 
bosonic system. 
It consists of $N_I=1,2$ non-interacting bosonic impurities (I) possessing a $1/2$ pseudospin  \cite{Kasamatsu}, immersed in a structureless bosonic medium (B) with $N_B=100$ atoms. 
The impurity-bath system here is mass balanced, $m_B=m_I$ (unless stated otherwise), while the involved components 
are harmonically confined in the same 1D external potential of frequency $\omega$. 
This scenario can be realized, for instance, by assuming a $^{87}$Rb gas. 
Here, the impurity pseudospin degree of freedom corresponds e.g. to the hyperfine states $\ket{\uparrow}\equiv\ket{F=1,m_F=1}$ and 
$\ket{\downarrow}\equiv\ket{F=1,m_F=-1}$ whilst the bath bosons are in the $\ket{F=2,m_F=1}$ state \cite{Egorov,Alvarez,Katsimiga_DAD}. 
Another experimentally relevant setting is, to a good approximation, a mixture of isotopes containing two hyperfine states of $^{85}$Rb 
for the impurity subsystem and of $^{87}$Rb for the bosonic bath. 

The resulting MB Hamiltonian reads
\begin{equation}
\begin{split}
\hat{H} = \hat{H}^{0}_{B}+\hat{H}_{BB}+\sum\limits_{a=\uparrow, \downarrow} \hat{H}^{0}_a 
+\hat{H}_{BI}+\hat{H}_S.
\label{Htot_system}
\end{split}
\end{equation} 
Here, $\hat{H}^{0}_{B}=\int dx~\hat{\Psi}^{\dagger}_{B} (x) \left( -\frac{\hbar^2}{2 m_B} 
\frac{d^2}{dx^2} +\frac{1}{2} m_B \omega^2 x^2 \right) \hat{\Psi}_{B}(x)$, 
and $\hat{H}^{0}_a=\int dx~\hat{\Psi}^{\dagger}_a(x) \left( -\frac{\hbar^2}{2 m_I} \frac{d^2}{dx^2}  
+\frac{1}{2} m_I \omega^2 x^2 \right) \hat{\Psi}_a(x)$ denote the non-interacting 
Hamiltonian of the BEC and the impurity respectively. 
The field operator $\hat{\Psi}_{B} (x)$ refers to the BEC while $\hat{\Psi}_{a} (x)$ to the 
spin-$a$ impurity atoms with $a=\left\{ \uparrow, \downarrow \right \}$. 
We further consider that the dominant interparticle interaction is an $s$-wave one since we operate 
in the ultracold regime~\cite{Olshanii}. 
As such both intra- ($g_{BB}$) and inter-component ($g_{BI}$) interactions are adequately 
described by contact ones. 
Moreover, $\hat{H}_{BB}=(g_{BB}/2) \int dx~\hat{\Psi}^{\dagger}_{B}(x) \hat{\Psi}^{\dagger}_{B}(x) 
\hat{\Psi}_{B} (x)\hat{\Psi}_{B}
(x)$ corresponds to the contact intracomponent interaction term of the bosonic bath. 
Also, in order to mimic the rf scheme only the spin-$\uparrow$ component of the impurities couples with the BEC and thus 
$g_{B \uparrow} \equiv g_{BI}$, while the spin-$\downarrow$ one is non-interacting namely $g_{B \downarrow}=0$. 
Additionally, the relevant intercomponent interaction term reads 
$\hat{H}_{BI}=g_{BI}\int dx~\hat{\Psi}^{\dagger}_{B}(x) \hat{\Psi}^{\dagger}_{\uparrow}(x) 
\hat{\Psi}_{\uparrow}(x)\hat{\Psi}_{B}(x)$. 

It is worth mentioning that the three-dimensional $s$-wave scattering length, ${a^s_{\sigma \sigma'}}$, for intra- or inter-species interactions, $\sigma=B,\uparrow,\downarrow$ is related to the effective one-dimensional coupling 
strength~\cite{Olshanii}. 
As a consequence $g_{\sigma\sigma'}$, is experimentally tunable either by adjusting ${a^s_{\sigma \sigma'}}$ using Feshbach 
resonances \cite{Kohler,Chin} or by means of the transversal confinement frequency ${{\omega _ \bot }}$ with the aid of 
confinement-induced resonances~\cite{Olshanii}. 

The impurity spin Hamiltonian with Rabi coupling is, 
\begin{equation}
\hat{H}_S=\frac{\hbar \Omega_{R0}(t)}{2} \hat{S}_x + \frac{\hbar \Delta}{2} \hat{S}_z, \label{spin_flip}
\end{equation}
with $\Omega_{R0} (t)$ and $\Delta=\nu -\nu_0$ referring to the time-dependent bare Rabi frequency and the detuning 
of the rf field~\cite{Jorgensen,Cetina} when the bosonic bath is absent. 
In particular, $\Omega_{R0} (t)$ and $\Delta$ characterize the intensity and frequency respectively of the applied rf pulse~\cite{Scully}. 
Notice also that the total spin operator is $\hat{\textbf{\textit{S}}}=\int dx~ \hat{\textbf{\textit{S}}}(x)=\int dx \sum_{ab} \hat{\Psi}_a^{\dagger} (x) 
\text{\boldmath$\sigma$}_{ab} \hat{\Psi}_b (x)$, with \text{\boldmath$\sigma$} denoting the Pauli 
vector~\cite{Tannoudji}. 

For convenience, in the following, the MB Hamiltonian of Eq.~(\ref{Htot_system}) is rescaled with 
respect to $\hbar  \omega$. 
Accordingly, the length, time, detuning and interaction strengths are expressed in units of $\alpha=\sqrt{\hbar/(m \omega)}$, $\omega^{-1}$, $\omega$ and $\hbar \omega \alpha =\sqrt{(\hbar^3 \omega)/m}$ respectively. 
A corresponding experimental realization of our one-dimensional setup e.g. when considering a $^{87}$Rb bosonic gas with 
$g_{BB}= 0.5 \sqrt{\hbar^3 \omega/m}\approx 3.55 \times 10^{-38}$ Jm can be easily achieved, for instance, by using a trap 
frequency $\omega=2\pi \times 100$ Hz while the transverse frequencies $\omega_{\perp}\approx 2\pi \times 5.1$ KHz. 
Here, temperature effects are suppressed for 
$k_BT \ll \frac{3^{4/3}}{16} (\frac{\alpha_{\perp}^2 N_B^2}{a_{BB}^s \alpha})^{2/3} \hbar \omega= 316 \hbar \omega \approx 1.5$~$\mu K$~\cite{Pethick_book}, 
where $k_B$ is the Boltzmann constant, $T$ is the temperature of the bosonic environment and $\alpha_{\perp}=\sqrt{\hbar /(m \omega_{\perp})}$ 
is the transversal confinement length scale. 

\begin{figure*}[ht]
\includegraphics[width=0.8\textwidth]{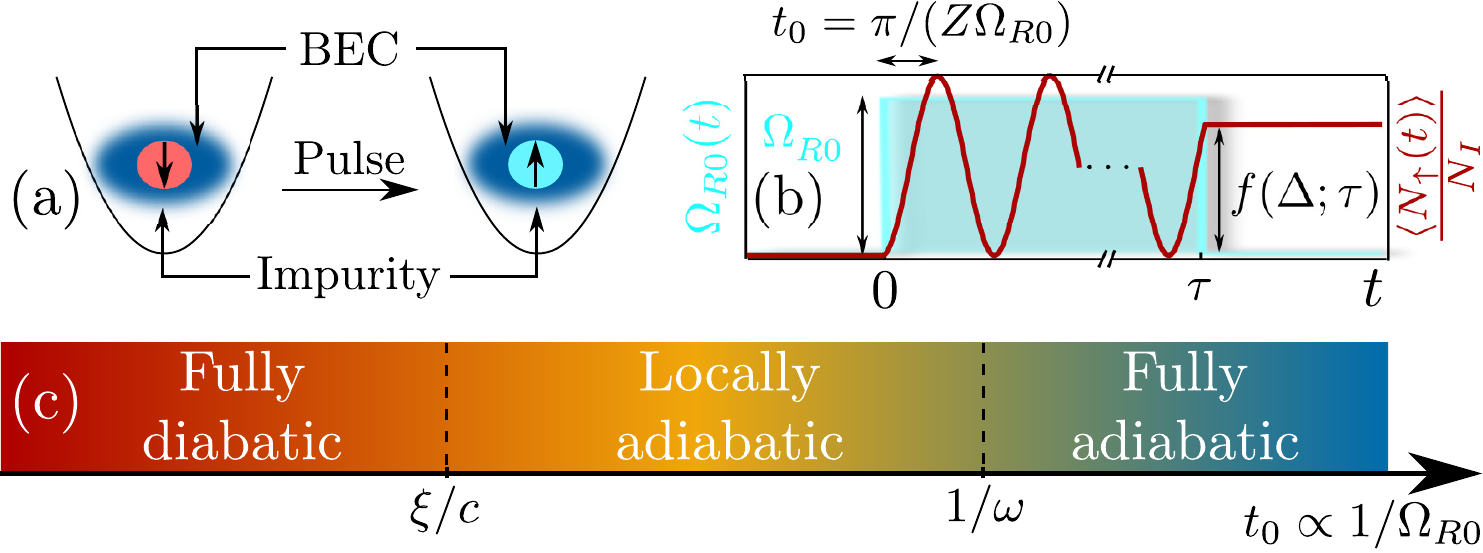}
\caption{(a) The impurity spin configuration before (left panel) and after (right panel) the application of the rectangular 
rf pulse. 
(b) Schematic representation of the utilized rectangular pulse (light-blue shaded region) imposed within the time-interval $0<t<\tau$ and the expected time-evolution of the population of the spin-$\uparrow$ 
atoms $\braket{N_{\uparrow}(t)}/N_I$ (red line) during the rf sequence. 
The spectroscopic signal $f(\Delta;\tau)$ is also illustrated [Eq.~(\ref{transf_fraction})]. 
(c) Categorization of the involved spin-transfer regimes as described by the characteristic timescale of the spin-dynamics $t_0$ in terms of the bare Rabi frequency $\Omega_{R0}$ (intensity) of 
the rf pulse. Here, $\omega$ denotes the external trapping frequency, $c$ is the speed of sound, $\xi$ the healing length of 
the BEC medium and $Z$ refers to the polaron residue.}
\label{fig:schematic} 
\end{figure*}

To trigger the nonequilibrium dynamics, see Fig.~\ref{fig:schematic} (a), we prepare the bosonic bath in its ground state  with $g_{BB}=0.5$ and the impurities in their spin-$\downarrow$ state being non-interacting with the bath. 
As a result the system is initialized in the $\ket{N_I/2,-N_I/2}=\bigotimes_{i=1}^{N_I}\ket{\downarrow}_i$ spin configuration where $\langle\hat{H}_{BI}\rangle=0$. 
The spinor part of the wave function is expressed in the basis of the total spin, namely $\ket{S,S_z}$~\cite{Tannoudji}. 
Moreover, the spatial part $\ket{\Psi^0_{BI}}$ of the system's ground state satisfies the eigenvalue equation 
$(\hat{H}-\hat{H}_{BI})\ket{\Psi^0_{BI}} \ket{N_I/2,-N_I/2}=E_0\ket{\Psi^0_{BI}}\ket{N_I/2,-N_I/2}$, with 
$\hat{H}_{BI}\ket{\Psi^0_{BI}} \ket{N_I/2,-N_I/2}=0$ and $E_0$ being the corresponding eigenenergy. 

For $0<t<\tau$, with $\tau$ denoting the exposure time, a rectangular pulse, 
$\Omega_{R0}(t)=\Omega_{R0}~\theta(t)~\theta(\tau-t)$ 
where $\theta$ is the heaviside theta function, 
induces Rabi oscillations between the spin-$\uparrow$ ($g_{BI}\neq0$) and spin-$\downarrow$ ($g_{BI}=0$) impurity 
states~\cite{Schmidt_rev} in the course of time $t$, see Fig.~\ref{fig:schematic} (b). 
In particular, for fixed $\Omega_{R0}$ the frequency of these Rabi oscillations depends on the detuning $\Delta$ following 
$\Omega_R=\sqrt{(Z\Omega_{R0})^2+(\Delta-\Delta_+)^2}$ where $\Delta_+$ is the resonance frequency associated with the 
Bose polaron formation characterized by a quasi-particle residue $Z$~\cite{Massignan,Schmidt_rev}. 
The latter refers to the overlap between the initial non-interacting state $\ket{\Psi_0}$ of the system and the polaronic state $\ket{\Psi_P}$ 
after a single spin-flip, i.e. $Z=\abs{\braket{\Psi_0|\hat{S}_x|\Psi_P}}$. 
As we shall demonstrate below, the Rabi oscillations are damped due to the accumulation
of spin-spatial correlations which are captured by the impurity-BEC complex. 
These correlations, in turn, lead to the decaying behavior of the polaron corresponding to $Z< 1$. 

To monitor the above-described effects for different exposure times $\tau$ and variable detuning $\Delta$, we 
resort to the fraction of impurity atoms that have been successfully transferred to the $\ket{\uparrow}$-state [Fig.~\ref{fig:schematic} (b)], namely 
\begin{equation}
f(\Delta;\tau)=\frac{\langle N_{\uparrow} (t=\tau)\rangle}{N_I},\label{transf_fraction} 
\end{equation}
where $N_I=N_{\uparrow}+N_{\downarrow}$. 
Due to the parabolic confinement, the timescale for the evolution of the spatial part of the MB wave function 
is given by $\sim \omega^{-1}$. 
The characteristic energy scale of the phononic excitations in the BEC is $\epsilon_{ph}=\hbar c /\xi$, 
where $c=\sqrt{g_{BB} n_B/m_B}$ denotes the speed of sound, $\xi=\hbar (2m_B g_{BB} n_B)^{-1/2}$ refers to the healing 
length and $n_B$ is the peak density. 
The relevant timescale for impurity-medium correlations to develop is then, $t_{phI}=\xi/c\approx 0.08 \omega^{-1}$. 
Moreover, the characteristic timescale for the evolution of the spin degrees of freedom is $t_0 \sim \Omega_{R0}^{-1}$, 
see Eq.~(\ref{spin_flip}) and Fig.~\ref{fig:schematic} (b). 
As a consequence, there is a competition between the characteristic times of the spatial and spin evolution which affects the emergent correlation properties. 

In particular, for $\Omega_{R0}\gg c/ \xi \gg \omega$ (herein $c/ \xi\approx 12.5$) the spatial MB wave function is unable to evolve during the short timescale, $T=2\pi/(Z\Omega_{R0})\approx t_0$), of the pulse-induced impurity Rabi oscillations. 
As such within this timescale, impurity-BEC correlations can hardly develop.  
This limiting case is referred to in the following as the {\it fully diabatic spin-transfer regime} [Fig.~\ref{fig:schematic} (c)]. 
When $\Omega_{R0}\ll \omega \ll c/ \xi$, on the other hand, the evolution of the spatial MB wave function is 
much faster than the 
impurity spin transfer, and we refer to a {\it fully adiabatic spin-transfer regime} [Fig.~\ref{fig:schematic} (c)]. 
Here, the introduction of the rf pulse results in the transition from non-interacting (for spin-$\downarrow$ impurities) 
to interacting (for spin-$\uparrow$ impurities) eigenstates. 
Notice that the latter states are not necessarily the polaronic states e.g. due to the TOC 
phenomenon~\cite{Mistakidis_orth_cat,Mistakidis_pump}. 

Remarkably, within the intermediate spin-transfer regime characterized by $\omega<\Omega_{R0}< c/ \xi$, the spatial correlations can build-up while effects due to the external confinement are suppressed. 
For these reasons, in the present work we are mainly interested in this intermediate intensity regime of the rf pulse which essentially leads to 
a {\it locally adiabatic} transfer of the impurities where the above-mentioned competition of timescales as defined by the trap ($\omega^{-1}$) and the phononic excitations ($\xi/c$) is pronounced. 
The resultant polaron formation in the latter regime will be contrasted to the one emanating when approaching the diabatic limit for 
increasing $\Omega_{R0}$. 

\subsection{Variational treatment of the spectral response} \label{sec:variational_ansatz} 

To address the rf spectroscopy of the above-discussed particle imbalanced multicomponent bosonic system we solve the 
underlying time-dependent MB Schr{\"o}dinger equation by resorting to the ML-MCTDHX method~\cite{MLX,MLB1}. 
It constitutes a variational approach~\cite{Lode_review} for simulating the nonequilibrium quantum dynamics of bosonic and fermionic 
multicomponent spinor settings~\cite{mistakidis_phase_sep,Mistakidis_eff_mass,Mistakidis_orth_cat,Mistakidis_Fermi_pol}. 
More precisely, in order to capture the emergent impurity-BEC correlations, we first express the total MB 
wave function $|\Psi(t)\rangle$ as a linear combination of $k=1,2,\dots,D$ distinct orthonormal species functions, i.e. 
$|\Psi^{\sigma}_k(t)\rangle$~\cite{MLX} for each individual component $\sigma=B,I$.  
This procedure results to a truncated Schmidt decomposition~\cite{Horodecki}, possessing a rank $D$, which reads     
\begin{equation} 
|\Psi(t)\rangle=\sum_{k=1}^D \sqrt{\lambda_k(t)} |\Psi^{\rm B}_k(t)\rangle|\Psi^{\rm I}_k(t)\rangle.  
\label{eq:wfn}
\end{equation} 
The involved time-dependent expansion coefficients $\lambda_k(t)$ are termed Schmidt weights and are also known as 
the natural populations of the $k$-th species function. 
In particular, they correspond to the eigenvalues of the $\sigma$-component reduced density matrix i.e. 
$\rho_{\sigma}^{N_{\sigma}} (\vec{x}, \vec{x}';t)= \braket{\Psi(t)|\prod_{i=1}^{N_{\sigma}} \Psi_{\sigma}^{\dagger} 
 (x_i) \prod_{i=1}^{N_{\sigma}} \Psi_{\sigma} 
 (x_i')|\Psi(t)} $, with $\vec{x}=(x_1, \cdots,x_{N_{\sigma}})$. 
 By inspecting Eq.~(\ref{eq:wfn}) it becomes also apparent that the system is entangled~\cite{Roncaglia}, and thus 
impurity-medium correlations are finite, when at least two different $\lambda_k(t)$ are nonzero; otherwise 
$|\Psi(t)\rangle$ is a direct product of two states and the system is non-entangled. 

Subsequently, for incorporating the intracomponent correlations into the MB ansatz, each species function is 
expanded in terms of a time-dependent number-state basis set $|\vec{n} (\tau) \rangle^{\sigma}$ as follows  
\begin{equation}
    | \Psi_i^{\sigma} (t) \rangle =\sum_{\vec{n}} A^{\sigma}_{i;\vec{n}}(t) | \vec{n} (t) \rangle^{\sigma}.  
    \label{eq:number_states}
\end{equation} 
Here, $A^{\sigma}_{i;\vec{n}}(t)$ denote the corresponding time-dependent expansion coefficients, while a particular 
number state $|\vec{n} (t) \rangle^{\sigma}$ refers to a permanent building upon $d^{\sigma}$ time-dependent variationally 
optimized single-particle functions (SPFs). 
The latter read $\left|\phi_l^{\sigma} (t) \right\rangle$, $l=1,2,\dots,d^{\sigma}$ with 
$\vec{n}=(n_1,\dots,n_{d^{\sigma}})$ being the corresponding occupation numbers. 
Next, the SPFs are expressed on a time-independent primitive basis. 
For the bosonic medium this refers to an $\mathcal{M}$ dimensional discrete variable representation (DVR) being 
designated by $\lbrace \left| k \right\rangle \rbrace$. 
Turning to the impurities this basis is the tensor product of the DVR basis, $\lbrace \left| k,s \right\rangle \rbrace$, 
for the spatial degrees of freedom and the two-dimensional pseudospin-$1/2$ basis $\{\ket{\uparrow}, \ket{\downarrow}\}$. 
For our investigation $\mathcal{M}=800$ grid points of a sine DVR are utilized. 
Therefore, a specific impurity SPF is a spinor wave function of the form  
\begin{equation}
| \phi^{\rm I}_j (t) \rangle= \sum_{k=1}^{\mathcal{M}}\big( B^{{\rm
I}}_{jk \uparrow}(t) \ket{k} \ket{\uparrow}+B^{{\rm I}}_{jk \downarrow}(t) \ket{k} \ket{\downarrow}\big).  
\label{eq:spfs}
\end{equation}
In this expression, the time-dependent expansion coefficients of the pseudospin-$\uparrow$ [$\downarrow$] are 
$B^{{\rm I}}_{j k \uparrow}(t)$ [$B^{{\rm I}}_{j k \downarrow}(t)$], see also 
Refs.~\cite{Mistakidis_orth_cat,Mistakidis_pump} for a more elaborate discussion. 

To determine the time-evolution of the underlying ($N_B+N_I$)-body wave function $\left|\Psi(t) \right\rangle$ 
subjected to the Hamiltonian of Eq.~(\ref{Htot_system}) the so-called ML-MCTDHX equations of motion~\cite{MLX} are 
solved numerically.  
These are obtained by following the Dirac-Frenkel variational principle~\cite{Frenkel,Dirac} for the ansatz exemplified in 
Eqs.~(\ref{eq:wfn}), (\ref{eq:number_states}) and (\ref{eq:spfs}) and results in a set of $D^2$ linear differential equations 
of motion regarding the $\lambda_k(t)$ coefficients being coupled to 
$D(\frac{(N_B+d^B-1)!}{N_B!(d^B-1)!}+\frac{(N_I+d^I-1)!}{N_I!(d^I-1)!})$ and $d^B+d^I$ nonlinear integrodifferential 
equations for the species functions and the SPFs respectively. 

Concluding, the main facet of this method is that it relies on the expansion of the system's MB wave function with respect to 
a time-dependent and variationally optimized basis which is accordingly tailored to account for all the important 
interparticle spatial and spin-spin correlations~\cite{Lode_review}. 
Simultaneously, the time-dependence of the basis enables for efficiently spanning the corresponding subspace of 
the Hilbert space at each time-instant for systems 
including mesoscopic particle numbers in different spatial dimensions~\cite{Katsimiga_DB,Axel_exp}. 
The Hilbert space truncation is signified by the used orbital configuration space, i.e. $C=(D;d^B;d^I)$. 
For the considered system the Bose gas comprises of a large number of weakly interacting atoms and as a consequence its 
intracomponent correlations are suppressed. 
Therefore, they can be captured by utilizing a small number of orbitals, here $d^B<4$. 
On the contrary, since $N_I=1,2$, the number of integrodifferential equations is small which enables us to employ many 
orbitals, here $d^I=D=8$, and thus account for strong impurity-impurity and impurity-medium correlations. 
Consequently, the amount of the underlying ML-MCTDHX equations of motion that should be evaluated is tractable.

\subsection{Approximate spectral methods}\label{sec:approximate_methods} 

\subsubsection{Effective potential}\label{sec:effect_pot}

To provide an intuitive understanding of the impurities spectral response and expose the necessity of including intra- and 
intercomponent correlations in the course of the evolution at least for certain intensities of the rf pulse  
we invoke an effective 
potential picture~\cite{Hannes,Theel} which has been shown to be adequate for $\abs{g_{BI}}<g_{BB}$~\cite{Mistakidis_orth_cat,Mistakidis_inject_imp,Mistakidis_induced_cor}. 
It is constructed by the external harmonic oscillator and the density of the bosonic background accounting also 
for the characteristics of the pulse i.e. the bare Rabi frequency $\Omega_{R0}$ and detuning $\Delta$. 
It essentially corresponds to a modified external harmonic potential and reads 
\begin{equation}
\begin{split}
&\hat{V}_{eff}(g_{BI})=\int dx~\bigg\{\bigg[ \frac{1}{2} m_I \omega^2 x^2 + \frac{g_{BI}}{2} \rho^{(1)}_B(x)\bigg] \hat{\mathbb{I}}(x) 
\\&+ \bigg( \frac {g_{BI}}{2} \rho^{(1)}_B(x)- \frac{\hbar \Delta}{2} \bigg) \hat{S}_z(x) +\frac{\hbar \Omega_{R0}}{2} \hat{S}_x(x) \bigg\} ,
\label{eq:eff_potential}
\end{split}
\end{equation}
where $\rho^{(1)}_B(x)\equiv\rho^{(1)}_B(x;g_{BI}=0)$ refers to the ground state density of the medium~\cite{comment1} for vanishing impurity-BEC coupling strength ($g_{BI}=0$)  and 
$\hat{\textbf{\textit{S}}}(x)=\sum_{ab} \hat{\Psi}_a^{\dagger}(x) 
\text{\boldmath$\sigma$}_{ab} \hat{\Psi}_b (x)$, $\hat{\mathbb{I}}(x)=\sum_{a} \hat{\Psi}_a^{\dagger} (x) \hat{\Psi}_a (x)$. 
As it can be readily seen, the form of $V_{eff}(x;g_{BI})$ implies that $\rho^{(1)}_B(x)$ acts on the impurities 
as an additional repulsive (for $g_{BI}>0$) or attractive (if $g_{BI}<0$) potential. 

It is also worth mentioning that $V_{eff}(x;g_{BI})$ neglects several phenomena that might be important for the description 
of the impurity problem. 
In particular, $V_{eff}(x;g_{BI})$ disregards the impurity-medium correlations. 
Consequently, it does not include the 
renormalization of the impurity mass caused by the coupling with its environment and the 
possible emergence of impurity-impurity induced interactions due to the absence of two-body terms. 
Thus, it is naturally expected that two-body processes can not be captured within $V_{eff}(x;g_{BI})$. 
To track the nonequilibrium dynamics of the impurities within $V_{eff}(x;g_{BI})$ we solve the underlying 
time-dependent single-particle Schr{\"o}dinger equation.

\subsubsection{Diabatic approximation}\label{sec:diab_approx}

Another interesting approximation that we shall employ in the following in order to unravel the impurity 
injection spectrum and importantly elucidate the transition to the {\it diabatic spin-transfer regime} 
where $\Omega_{R0}\gg c/ \xi$ is the so-called diabatic approximation. 
As exemplified in Sec.~\ref{sec:hamiltonian_protocol}, in this regime the spatial part of the MB
wave function is not possible to adapt to the external perturbation within the small timescale of the pulse and 
hence the emergent impurity-medium correlations are expected to vanish. 
The main ingredient of the diabatic approximation is that in the course of the pulse the kinetic energy of the 
BEC and the impurities is considered to be negligible, and thus we treat the participating atoms as stationary 
ones during the application of the injection pulse. 
This can be understood by the fact that in the case of $\Omega_{R0} \gg c/ \xi$ the impurities Rabi oscillation 
between their spin-$\downarrow$ and spin-$\uparrow$ states takes place at a much faster timescale 
i.e. $\sim1/\Omega_{R0}$ as compared to the relative motion among the bath and the impurity atoms which is 
of the order of $\sim \xi/c$. 
The latter constitutes the characteristic timescale in which elementary BEC excitations form. 

It can be systematically shown [see for details Appendix~\ref{app:diabatic_approx}] that within the diabatic 
approximation ($\Omega_{R0} \gg c/ \xi> \omega$) the time-averaged rf signal, employing a second-order 
Taylor expansion in terms of $\Delta/\Omega_{R0}$, has the following form   
\begin{equation}
\begin{split}
\bar{f}(\Delta)&=\frac{1}{2} \frac{\tilde{\Omega}^2}{\tilde{\Omega}^2+\tilde{\Delta}^2} \bigg[
1 +\frac{\tilde{\Delta}}{\tilde{\Omega}^2+\tilde{\Delta}^2} + \frac{3}{8} 
\frac{3 \tilde{\Delta}^2 - \tilde{\Omega}^2}{( \tilde{\Omega}^2 +\tilde{\Delta}^2 )^2} 
\\& +\mathcal{O} \left(\frac{\tilde{\Delta}^3}{( \tilde{\Omega}^2 +\tilde{\Delta}^2)^3},
\frac{\tilde{\Omega}^3}{( \tilde{\Omega}^2 +\tilde{\Delta}^2)^3} \right) 
\bigg].
\label{eq:diabatic_spec}
\end{split}
\end{equation}
Here, $\tilde{\Omega}(\Omega_{R0})=\frac{2 g_{BB} \Omega_{R0}}{g_{BI} \omega}$ and 
$\tilde{\Delta}(\Delta)=\frac{2g_{BB}}{g_{BI}} [ \frac{\Delta}{\omega} + \frac{1}{2} \frac{g_{BI}}{g_{BB}} (\frac{R_{TF}}{\alpha})^2]$ 
where $\alpha=\sqrt{\hbar/(m \omega)}$ is the harmonic oscillator length and $R_{TF}$ is the Thomas-Fermi radius of the BEC background. 
It is worth mentioning that Eq.~(\ref{eq:diabatic_spec}) implies that in the {\it diabatic spin-transfer regime} the 
impurity spectrum corresponds to a Lorentzian. 
The latter is centered at $\tilde{\Delta}(\Delta_{0}) \approx - 0.5$, or equivalently at 
$\Delta_0 \approx - \frac{\omega}{2} \frac{g_{BI}}{g_{BB}}\left[\left(\frac{R_{TF}}{\alpha} \right)^2 - \frac{1}{2} \right]$, 
and exhibits a half-width-at-half-maximum around $\gamma\approx \Omega_{R0}$. 
Importantly, we can readily deduce the linear dependence of the width of the peak $\gamma$ on the used bare Rabi 
frequency $\Omega_{R0}$. 
The origin of this result stems from the substantial power broadening of the rf spectrum for such high pulse intensities.

\begin{figure*}[ht]
\includegraphics[width=0.8\textwidth]{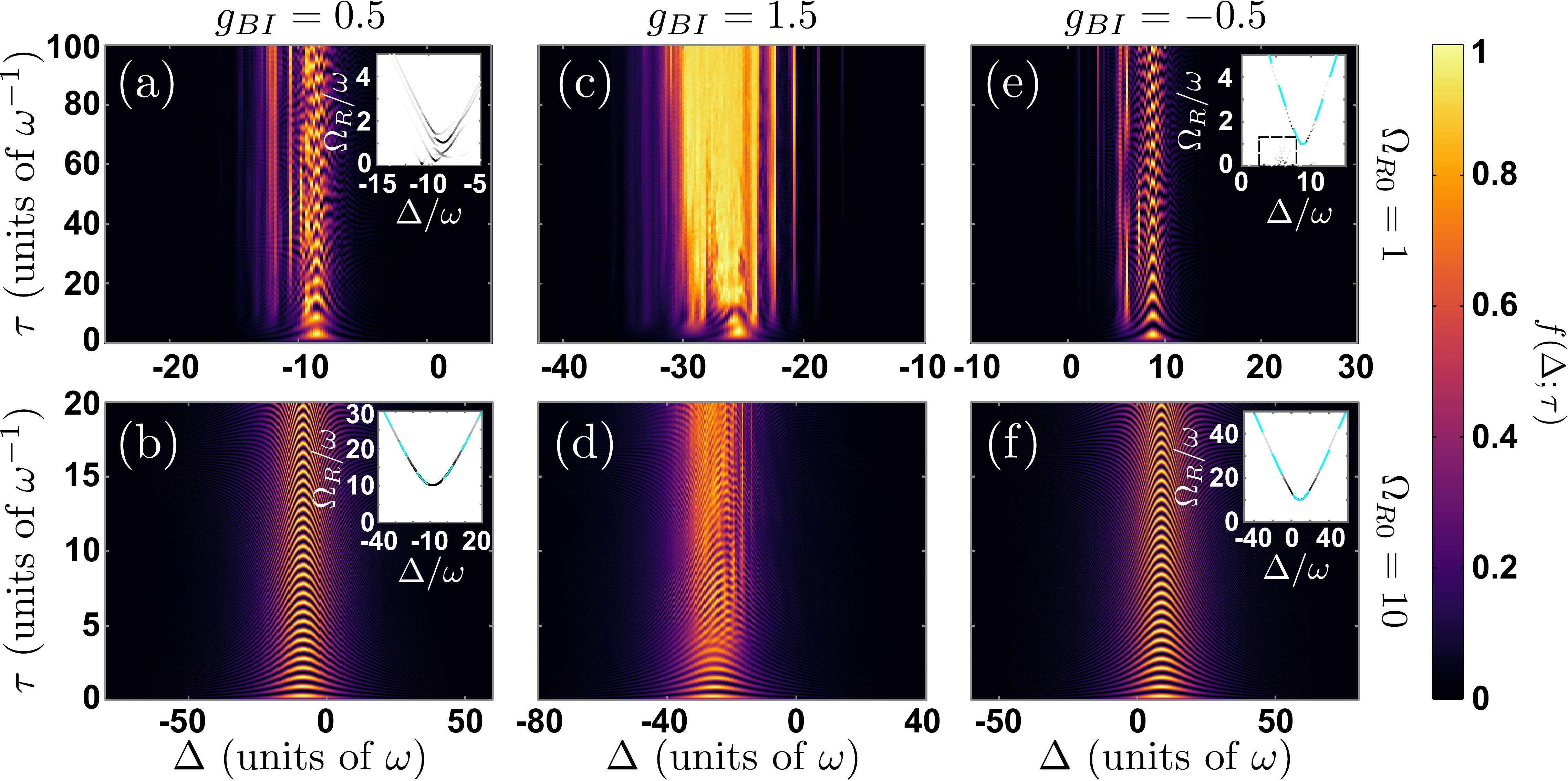}
\caption{Temporal evolution of the spectroscopic signal $f(\Delta;\tau)$ of a single impurity $N_I=1$, featuring multiple polaronic resonances, with respect to the 
detuning of the rf field and different impurity-bath couplings for a bare Rabi frequency (a), (c), (e) $\Omega_{R0}=1$ and (b), (d), (f) $\Omega_{R0}=10$. 
In particular, we consider (a), (b) $g_{BI}=0.5$, (c), (d) $g_{BI}=1.5$, and (e), (f) $g_{BI}=-0.5$ corresponding to the cases of repulsive Bose polaron, its dynamical decay (TOC) and the attractive polaron regimes respectively.    
The insets provide the corresponding Fourier spectrum of the signal $f(\Delta;\Omega_R)$. 
The dashed light-blue line in the insets of (b), (e), (f) constitutes a fitting of the theoretically anticipated 
$\Omega_R=\sqrt{(Z \Omega_{R0})^2+(\Delta-\Delta_+^*)^2}$, with 
$\Delta_+^*$ being the position of the resonance and $Z\approx 0.98$  [$Z\approx 0.99$] is the quasi-particle residue in (b) [(e), (f)]. The dashed box in the inset of (e) marks the presence of frequencies $\Omega_R$ corresponding to the spectral peaks at $\Delta<\Delta_+^*$. 
In all cases the mass-balanced ($m_B=m_I$) and harmonically trapped ($\omega=1$) system consists of $N_B=100$ bosons 
with $g_{BB}=0.5$ and a single impurity where $g_{B \downarrow}=0$. }
\label{fig:spectrum_weak_int} 
\end{figure*}

\section{Radiofrequency spectrum of impurities in a mass-balanced mixture}\label{sec:mass_balance}

To probe the impurity excitation spectrum we theoretically simulate the celebrated technique of reverse rf 
spectroscopy~\cite{Kohstall,Scazza,Gupta_rf}. 
The dynamical protocol characterizing this spectroscopic scheme consists of the following steps. 
Initially the multicomponent bosonic system satisfying the MB Hamiltonian of Eq.~(\ref{Htot_system}) is 
prepared in its ground state with zero impurity-medium interaction strength i.e. $g_{BI}=0$, a bare Rabi coupling 
$\Omega_{R0}$ and vanishing detuning $\Delta=0$. 
Particularly, the BEC medium of $N_B=100$ atoms resides in its interacting ground state with $g_{BB}=0.5$ 
and thus having a Thomas-Fermi radius $R_{TF}\approx 4.2$. 
The single as well as the two non-interacting bosonic impurities lie in their spin-$\downarrow$ (ground) 
state where $g_{B \downarrow}=0$. 

To trigger the impurity nonequilibrium dynamics a rectangular rf pulse of a specific bare Rabi frequency 
$\Omega_{R0}$ and varying detuning $\Delta$ for different realizations is subsequently applied, thus driving 
the impurity atoms to their 
interacting spin-$\uparrow$ state. 
Importantly, we study the resulting spectral response at specific impurity-medium couplings $g_{BI}$ 
for different $\Omega_{R0}$ with respect to $\Delta$ in 
order to understand the emergent correlation phenomena of the induced spin-flip dynamics and also categorize 
the latter at distinct spin-transfer regimes ranging from the adiabatic to the diabatic limit. 
More precisely, below, we focus on three different values of $g_{BI}$ being representative of the formation of repulsive ($g_{BI}=0.5$) and attractive ($g_{BI}=-0.5$) Bose polarons satisfying $\abs{g_{BI}} \leq g_{BB}$ as well as at 
strong repulsions ($g_{BI}=1.5$) i.e. $g_{BI}>g_{BB}$ yielding a dynamical decay of the emergent 
quasi-particle~\cite{Mistakidis_orth_cat}. 
The existence and stability of these states has been recently theoretically probed utilizing 
Ramsey~\cite{Mistakidis_induced_cor,Mistakidis_orth_cat} and pump-probe spectroscopy~\cite{Mistakidis_pump}. 

As explicated in Sec.~\ref{sec:hamiltonian_protocol} the simulated spectroscopic signal presented for instance 
in Fig.~\ref{fig:spectrum_weak_int} corresponds 
to the fraction of impurity atoms $f(\Delta;\tau)$ [Eq.~(\ref{transf_fraction})] that have been successfully transferred 
to the spin-$\uparrow$ state. 
The applied rectangular pulses [Fig.~\ref{fig:schematic} (b)] being abruptly switched on (off) at time $t=0$ ($t=\tau$) are characterized by fixed $\Omega_{R0}$, varying detuning $\Delta$ and pulse time $\tau$. 
The participating frequencies in $f(\Delta;\tau)$ are estimated by measuring the Fourier transform of the spectrum 
$f(\Delta;\Omega_R)=\int_0^T d\tau f(\Delta;\tau) e^{i \Omega_R \tau}$ with $T=\max (\tau)$ being the duration of 
the most extensive pulse. 
Moreover, in order to facilitate a comparison between the spectral responses at different intensities of the rf pulse and 
approximations we also invoke the time-averaged transfer fraction over $0<\tau<T$ being defined as 
$\bar{f}(\Delta)=(1/T)\int_0^T d\tau f(\Delta;\tau)$.  

To reveal the spatially resolved dynamics of the spin-$\uparrow$ and spin-$\downarrow$ impurity state as well as 
of the BEC medium we monitor the one-body reduced density matrix for each 
component~\cite{density_matrix}, namely 
\begin{equation}
\rho_\sigma^{(1)}(x,x';t)=\langle\Psi(t)|\hat{\Psi}_{\sigma}^{\dagger}(x)\hat{\Psi}_{\sigma}(x')
|\Psi(t)\rangle.\label{eq:single_par_den_matr}
\end{equation}
Here, $\hat{\Psi}_{\sigma}(x)$ is the $\sigma=B, \uparrow, \downarrow$-component bosonic field operator acting 
at position $x$ and obeying the standard bosonic commutation relations \cite{Pethick_book}. 
In the following, we will inspect the respective one-body densities of each 
component i.e.~$\rho_\sigma^{(1)}(x;t)\equiv\rho_\sigma^{(1)}(x,x'=x;t)$. 
Recall that this quantity is experimentally accessible via averaging over a sample of single-shot 
images \cite{Bloch_review,mistakidis_phase_sep}.

\subsection{Multimode dynamics of coherent repulsive Bose polaron}\label{sec:mode_int} 
\begin{figure*}[ht]
\includegraphics[width=0.9\textwidth]{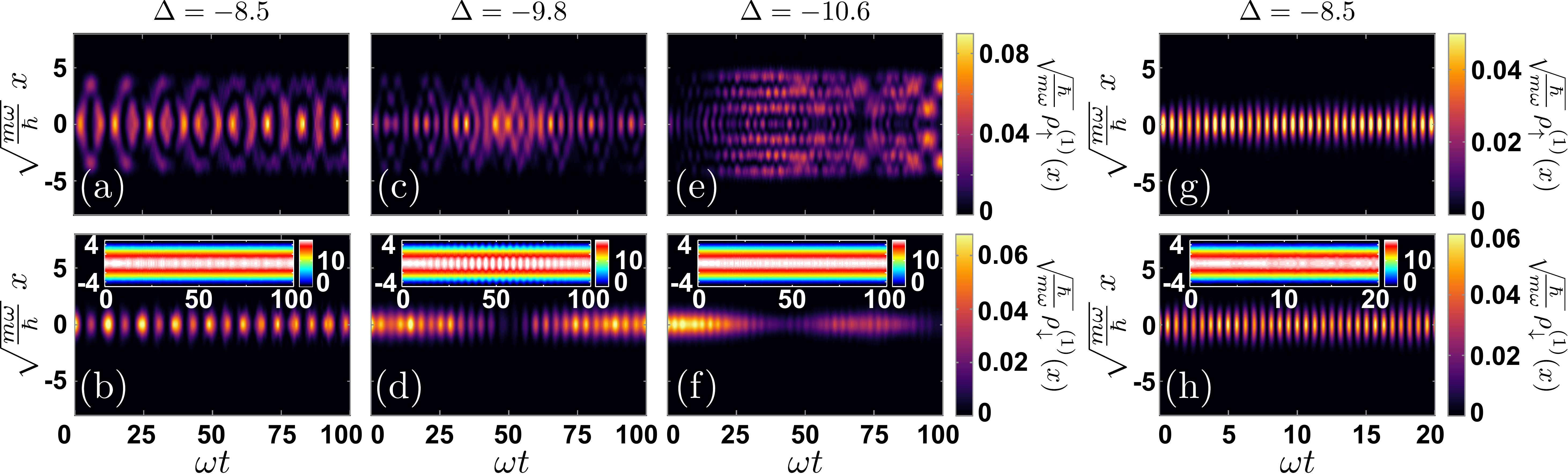} 
\caption{ Spatiotemporal evolution of the single-particle density (a), (c), (e) $\rho^{(1)}_{\uparrow}(x;t)$ (spin-$\uparrow$) 
and (b), (d), (f) $\rho^{(1)}_{\downarrow}(x;t)$ (spin-$\downarrow$) of a single impurity for different detunings 
$\Delta$ of the rf field (see legends) and fixed bare Rabi frequency $\Omega_{R0}=1$. 
The repulsive $g_{BI}$ leads to a delocalization trend of the impurity density indicating its excited nature.   
The insets of (b), (d), (f) present the corresponding density evolution $\rho^{(1)}_{B}(x;t)$ of the bosonic background where 
phononic excitations are imprinted. 
(g) $\rho^{(1)}_{\uparrow}(x;t)$ and (h) $\rho^{(1)}_{\downarrow}(x;t)$ of a single impurity for $\Delta=-8.5$ 
and $\Omega_{R0}=10$ featuring a more localized behavior than $\Omega_{R0}=1$ due to the faster spin dynamics. 
The respective $\rho^{(1)}_{B}(x;t)$ is showcased in the inset of (h). 
In all cases $g_{B \downarrow}=0$ and $g_{B\uparrow}\equiv g_{BI}=0.5$. 
The remaining system parameters are the same as in Fig.~\ref{fig:spectrum_weak_int}. 
The pulse time in (a)-(f) $\tau=T=100>t$ while in (g), (h) $\tau=T=20>t$. 
The Thomas-Fermi radius of the BEC medium is $R_{TF}\approx 4.2$.}
\label{fig:density_weak_int_weak_dr} 
\end{figure*}

Considering a bare Rabi frequency $\Omega_{R0}=\omega=1 \ll c/ \xi$ and for fixed repulsive impurity-BEC interactions $g_{BI}=0.5$, 
the obtained rf signal with varying $\Delta$ and pulse time $t$ is showcased in Fig.~\ref{fig:spectrum_weak_int} (a). 
Evidently, a resonance occurs for $\Delta_+^*=-8.5$ having three predominant Rabi frequencies, i.e. $\Omega_R \approx 1.05$, 
$\Omega_R \approx 0.63$ and $\Omega_R \approx 0.44$, see the inset of Fig.~\ref{fig:spectrum_weak_int} (a). 
The former signifies the existence of a repulsive Bose polaron~\cite{Ardila_MC,Mistakidis_induced_cor,Massignan}.
Notice here that $V_{eff}(x;g_{BI})$ for the spin-$\uparrow$ impurity is proximal to a square well of length 
$2R_{TF}$~\cite{Mistakidis_induced_cor}. 
Therefore, its lowest lying states possess energies $\epsilon_n\approx \epsilon_0+n^2 \pi^2\hbar^2/[2m_I(2R_{TF})^2]\approx 
\epsilon_0+0.0554\hbar\omega n^2$ with $n$ indexing each eigenstate. 
This implies that the involved energy differences are much smaller than $\Omega_{R0}=1$ 
and as a consequence the pulse acts almost diabatically on the impurity forming a superposition of the above-mentioned states. 
Indeed, monitoring the impurity spin densities we observe that $\rho_{\uparrow}^{(1)}(x;t)$ features a double and 
three peaked structure during the dynamics while expanding and contracting [Fig.~\ref{fig:density_weak_int_weak_dr} (a)] 
within its bosonic environment. 
Moreover, a multimode spin-transfer process between the $\ket{\downarrow}$ and $\ket{\uparrow}$ states occurs, containing the 
previously identified frequencies, due to the complex expansion dynamics of the spin-$\uparrow$ impurity portion. 
This multifrequency character is also imprinted in the dynamics of $\rho_{\downarrow}^{(1)}(x;t)$ majorly residing 
at $x\approx 0$ [Fig.~\ref{fig:density_weak_int_weak_dr} (b)]. 
The respective medium only shows marginal distortions, see the inset of Fig.~\ref{fig:density_weak_int_weak_dr} (a), 
referring to weakly excited phononic modes stemming from the polaron formation and dynamics. 

\begin{figure*}[ht]
\includegraphics[width=0.8\textwidth]{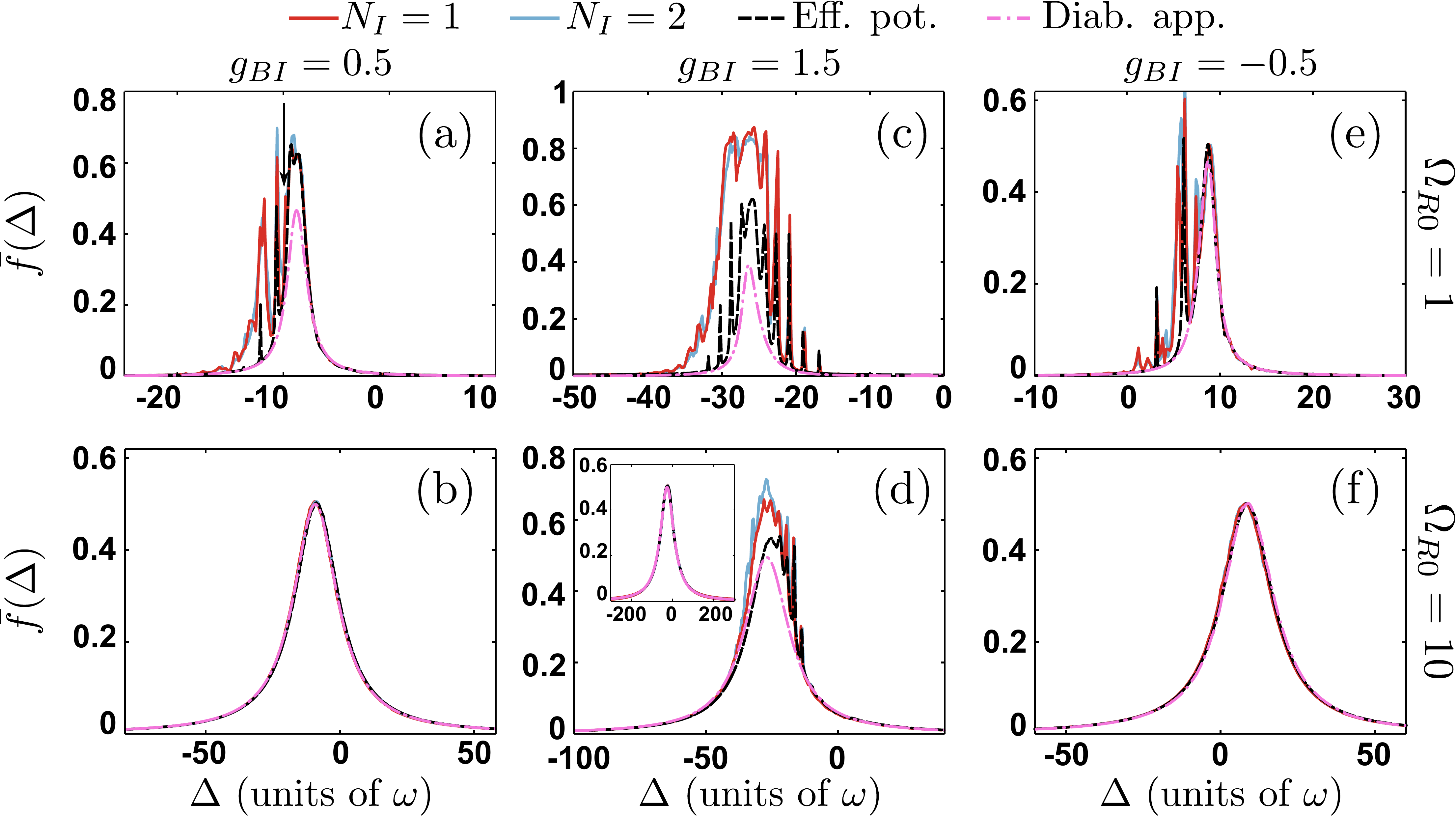}
\caption{Time-averaged spectroscopic signal $\bar{f}(\Delta)$ of a single ($N_I=1$) and two ($N_I=2$) bosonic non-interacting ($g_{\uparrow \uparrow}=0$) impurities and within different 
approaches (see legend). 
The cases of adiabatic (a), (c), (e) $\Omega_{R0}=1$ and diabatic (b), (d), (f) $\Omega_{R0}=10$ spin-transfer regimes are depicted for interactions resulting to the formation of (a), (b) repulsive $g_{BI}=0.5$, (e), (f) attractive $g_{BI}=-0.5$ and (c), (d) the dynamical decay $g_{BI}=1.5$ of the Bose polaron. 
The inset in (d) shows $\bar{f}(\Delta)$ for $\Omega_{R0}=40$.
The arrow in (a) indicates the spectral peak referring to the mode-coupling of the polaron with the collective 
background excitations. 
Other system parameters are the same as in Fig.~\ref{fig:spectrum_weak_int}.}
\label{fig:spectrum_weak_int_av} 
\end{figure*}

Other resonances taking place at smaller detunings, e.g. for $\Delta_+\approx -9.2$, $\Delta_+\approx-10.6$ 
and $\Delta_+\approx-12$, correspond to polaronic states with an excited motional degree-of-freedom, see also Fig.~\ref{fig:spectrum_weak_int_av} (a). 
Indeed, for a trapped system these states have a discrete (quantized) spectrum in contrast to the homogeneous case 
where the momentum of the polaron is a real-valued good quantum number~\cite{Schmidt_rev,Shchadilova,Grusdt_approaches}. 
They exhibit a much lower $\Omega_R$ when compared to the 
$\Delta_+=-8.5$ case. 
This decrease of $\Omega_R$ is due to the smaller overlap, or equivalently residue, of the excited polaronic states 
with the initial non-interacting one and it can be also be inferred by the slow spin-transfer dynamics 
[Figs.~\ref{fig:density_weak_int_weak_dr} (e), (f)]. 
In particular, these states are related to superpositions of energetically higher-lying eigenstates of $V_{eff}(x;g_{BI})$ 
introduced in Sec.~\ref{sec:effect_pot}, see also the discussion in Refs.~\cite{Mistakidis_orth_cat,Mistakidis_induced_cor}. 
Indeed, by inspecting the time-averaged spectrum $\bar{f}(\Delta)$ [Fig.~\ref{fig:spectrum_weak_int_av} (a)] 
the locations of the above-discussed resonant peaks are adequately captured by $V_{eff}(x;g_{BI})$ [Eq.~(\ref{eq:eff_potential})]. 
A clear manifestation of the character of these states appears in the evolution of $\rho_{\uparrow}^{(1)}(x;t)$ shown in 
Fig.~\ref{fig:density_weak_int_weak_dr} (e) for $\Delta_+=-10.6$. 
Evidently, $\rho_{\uparrow}^{(1)}(x;t)$ exhibits a multihump structure, indicative of the higher-lying excitations 
participating in the dynamics at these detunings. 
Additionally, the bosonic bath [inset of Fig.~\ref{fig:density_weak_int_weak_dr} (f)] is only slightly perturbed due to the 
sound wave emission associated with the polaronic excitations~\cite{Mistakidis_inject_imp,Mukherjee_pulse}. 

The remaining resonant peaks identified in the spectrum and are not present within 
$V_{eff}(x;g_{BI})$ [Fig.~\ref{fig:spectrum_weak_int_av} (a)], for 
instance the peak at $\Delta_+=-9.8$ [see the arrow in Fig.~\ref{fig:spectrum_weak_int_av} (a)], occur as a result of the 
coupling of the polaron resonances with the collective 
background excitations~\cite{Dolgirev,Mistakidis_induced_cor}. 
The latter refer majorly here to a breathing mode, see in particular the dynamics of $\rho_{B}^{(1)}(x;t)$ 
in the inset of Fig.~\ref{fig:density_weak_int_weak_dr} (d). 
As expected, an underlying spin-mixing dynamics among the impurity spin components, 
see Figs.~\ref{fig:density_weak_int_weak_dr} (c), (d), takes place with the $\rho_{\uparrow}^{(1)}(x;t)$ 
spreading within the BEC background and having a multinodal shape. 
Interestingly, within the time-interval ($25<t<75$) where the $\ket{\uparrow}$ state is maximally occupied the bosonic medium 
undergoes a pronounced breathing mode. 

Moreover, by comparing $\bar{f}(\Delta)$ between the MB scenario and the diabatic approximation, see 
Fig.~\ref{fig:spectrum_weak_int_av} (a), we can deduce that for $\Omega_{R0}=1$ the pulse intensity is not sufficient to reach the diabatic spin-transfer regime since the two spectra differ substantially. 
Interestingly, $\bar{f}(\Delta)$ as obtained in the MB approach for two non-interacting bosonic impurities 
almost coincides with the one of $N_I=1$ [Fig.~\ref{fig:spectrum_weak_int_av} (a)]. 
The latter suggests that here the emergent polaronic peaks are predominantly of single-particle character, 
a result that will be further analyzed and differences will be exposed later on in Sec.~\ref{sec:two_body}. 
However, by merely inspecting $\bar{f}(\Delta)$ the most essential difference between the $N_I=2$ and $N_I=1$ 
impurities is that in the former case the amplitude of the higher-lying peak at $\Delta_+\approx -10.6$ is larger. 
This alteration is caused by the larger number of impurities where the amount of excitations is enhanced. 

Turning to a more intense rf pulse with $\omega<\Omega_{R0}=10< c / \xi \approx 12.5$ (intermediate spin-transfer regime, Fig.~\ref{fig:schematic} (c)) and for $g_{BI}=0.5$ it becomes apparent that 
the spectral response shows a central polaronic resonance around $\Delta_+^*\approx -9.2$ [Fig.~\ref{fig:spectrum_weak_int} (b)]. 
It is also characterized by a single Rabi frequency which satisfies the relation 
$\Omega_R(\Delta)=\sqrt{(Z\Omega_{R0})^2+(\Delta-\Delta_+^*)^2}$ with respect to $\Delta$, see the inset of 
Fig.~\ref{fig:spectrum_weak_int} (b) where $Z\approx 0.98$. 
The impurity simply Rabi oscillates between its spin-$\downarrow$ and spin-$\uparrow$ states as it can easily 
inferred by monitoring 
its spin densities evolution presented in Fig.~\ref{fig:density_weak_int_weak_dr} (g), (h) exemplarily for $\Delta=\Delta_+^*$. 
Note that weak amplitude sound waves are accordingly imprinted onto the density of the BEC 
background~\cite{Mukherjee_pulse,Marchukov_sw} caused by the impurity 
motion in the latter [inset of Fig.~\ref{fig:density_weak_int_weak_dr} (h)]. 
Furthermore, the respective time-averaged signal $\bar{f}(\Delta)$ has the form of a Lorentzian distribution for both one and 
two bosonic impurities, a result that is further supported by its almost perfect agreement with $\bar{f}(\Delta)$ as predicted 
by the diabatic approximation [Fig.~\ref{fig:spectrum_weak_int_av} (b)]. 
The latter implies that within the dynamics for $ \Omega_{R0}=10$ the diabatic spin-transfer regime has already been reached, despite the fact that $\Omega_{R0}<c/ \xi$. 
Also, since $\bar{f}(\Delta)$ is unaltered for $N_I=1$ and $N_I=2$ it means that two-body processes are not manifest. 
In this limit the effective potential leads to exactly the same shape of $\bar{f}(\Delta)$, suggesting that impurity-BEC correlations are suppressed. 

The crossover from the adiabatic to the diabatic spin-transfer regime imprinted in the rf signal can be visualized in terms of the structural change of $\bar{f}(\Delta)$ for increasing $\Omega_{R0}$, 
see Fig.~\ref{fig:spectrum_weak_int_av_dif_freq} (a). 
Evidently, for $\Omega_{R0}\leq \omega$ the averaged spectrum $\bar{f}(\Delta)$ features a main polaronic resonance 
e.g. at $\Delta_+^* \approx -8.45$ [$\Delta_+^*\approx -8.5$] 
while secondary ones signifying higher-order polaronic excitations appear at smaller detunings for instance 
at $\Delta_+\approx -8.8$, $\Delta_+ \approx 9.65$ 
[$\Delta_+\approx -10.6$, $\Delta_+\approx -12$] for $\Omega_{R0}=0.1$ [$\Omega_{R0}=1$]. 
We remark that the number of these secondary resonances is smaller for decreasing $\Omega_{R0}\leq \omega$ 
since in the latter case 
the pulse is less intense and thus prohibits the spectral resolution of the higher-lying motional states 
of the polaron possessing $Z \ll 1$. 
Recall that the resonance peak for a rectangular pulse when $\Omega_{R0}\ll \omega$ is 
$f(\Delta_+;\tau)\sim \sin^2(Z\Omega_{R0} \tau/2)$~\cite{Mistakidis_pump} and thus low residue peaks cannot manifest within the timescale considered herein. 
In sharp contrast if $\omega \ll \Omega_{R0} < c/\xi$ even though the locally adiabatic regime is expected [Fig.~\ref{fig:schematic} (c)] the system exhibits a fully diabatic behavior as captured by $\bar{f}(\Delta)$ [Fig.~\ref{fig:spectrum_weak_int_av} (b)]. 
Indeed, the latter possesses a 
Lorentzian distribution whose width is enhanced for a larger $\Omega_{R0}$ due to power broadening, see 
also Eq.~(\ref{eq:diabatic_spec}). 
Here, $\bar{f}(\Delta)$ shows a single polaronic resonance with an $\Omega_{R0}$-dependent location for fixed $g_{BI}$. The above indicate that the impurity-bath coupling is adequately weak and therefore the impurity-phonon correlations can not affect the spin dynamics triggered by such intense pulses.

\subsection{Emergence of temporal orthogonality catastrophe}\label{sec:strong_int}

We now discuss the emergent rf response for strong repulsive coupling, and in particular for $g_{BI}=1.5>g_{BB}$. 
The rf signal $f(\Delta;\tau)$ upon applying a pulse with $\Omega_{R0}=\omega=1 \ll c/\xi $ and varying detuning is 
demonstrated in Fig.~\ref{fig:spectrum_weak_int} (c). 
A multitude of spectral peaks is observed in $f(\Delta;\tau)$ with the latter featuring a saturation tendency during 
the dynamics and in particular for $\tau>40$. 
This saturation trend together with the appearance of the multiple peaks in $f(\Delta;\tau)$ are clear manifestations of the dynamical decay, and consequently vanishing residue, of the emergent 
quasi-particle and its associated TOC phenomenon~\cite{Mistakidis_orth_cat,Mistakidis_pump}. 
The latter is caused by the build up of strong impurity-BEC correlations leading to the impurity dephasing
and phase separation from the bosonic medium, as we shall show below. 
\begin{figure*}[ht]
\includegraphics[width=0.8\textwidth]{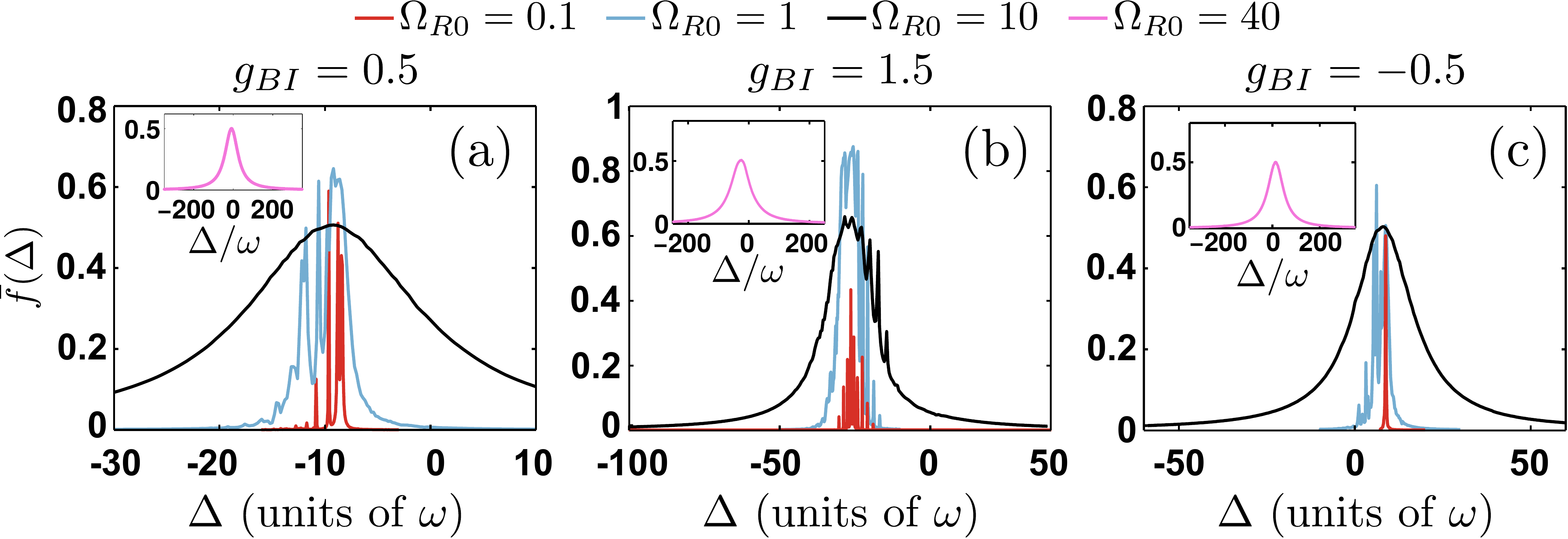}
\caption{Time-averaged spectroscopic signal $\bar{f}(\Delta)$ for varying detuning $\Delta$ and different impurity-medium interactions $g_{BI}$ as well as bare Rabi frequencies $\Omega_{R0}$ (see legends) in 
the MB case. 
The insets represent the $\bar{f}(\Delta)$ for $\Omega_{R0}=40$.
The crossover from adiabatic to diabatic spin-transfer is realized for an increasing $\Omega_{R0}$ and fixed $g_{BI}$ being captured by the modification of 
$\bar{f}(\Delta)$ from a multiple peaked structure to a Lorentzian shape. 
The remaining system parameters are the same as in Fig.~\ref{fig:spectrum_weak_int}.}
\label{fig:spectrum_weak_int_av_dif_freq} 
\end{figure*} 

The time-averaged spectrum $\bar{f}(\Delta)$ shown in Fig.~\ref{fig:spectrum_weak_int_av} (c) 
reveals the appearance of 
several resolved narrow peaks for $\Delta>-25$. 
Additionally a broad, i.e. possessing a width much larger than 
$\Omega_{R0}=1$, continuous distribution of excitations for $\Delta<-25$ occurs. 
The nature of these spectral features can be identified by contrasting the $\bar{f}(\Delta)$ for $N_I=1$ with the
corresponding one obtained within the effective potential approximation. 
Notice that the dynamics within $V_{eff}(x;g_{BI})$ cannot 
capture the phenomenon of TOC and is characterized by coherent evolution. 
Within this approximation, the spectrum is dominated 
by a prominent and power broadened peak at $\Delta_+^* \approx -25.8$, with additional individual 
narrower resonances for both larger and smaller detuning. 
For such strongly repulsive impurity-bose coupling, the potential $V_{eff}(x;g_{BI})$ for the spin-$\uparrow$ impurities possesses a 
double-well structure. 
The eigenstates corresponding to $\Delta_+^* \approx -25.8$ possess eigenenergies of the order of the barrier height and accordingly 
have a prominent single-particle density within the spatial extent of the barrier, i.e. within the BEC~\cite{Mistakidis_induced_cor}. 
Similarly to the case of $g_{BI}=0.5$, the energy levels of those states are closely packed and therefore the pulse acts almost diabatically creating a superposition. 
In contrast, the states of lower energy i.e. $\Delta > -25$ mainly reside in the periphery of the cloud at 
$x \approx R_{TF}$, while 
the states with $\Delta < -26$ show a highly multinodal structure implying the contribution of large impurity momenta. 

\begin{figure*}[ht]
\includegraphics[width=0.8\textwidth]{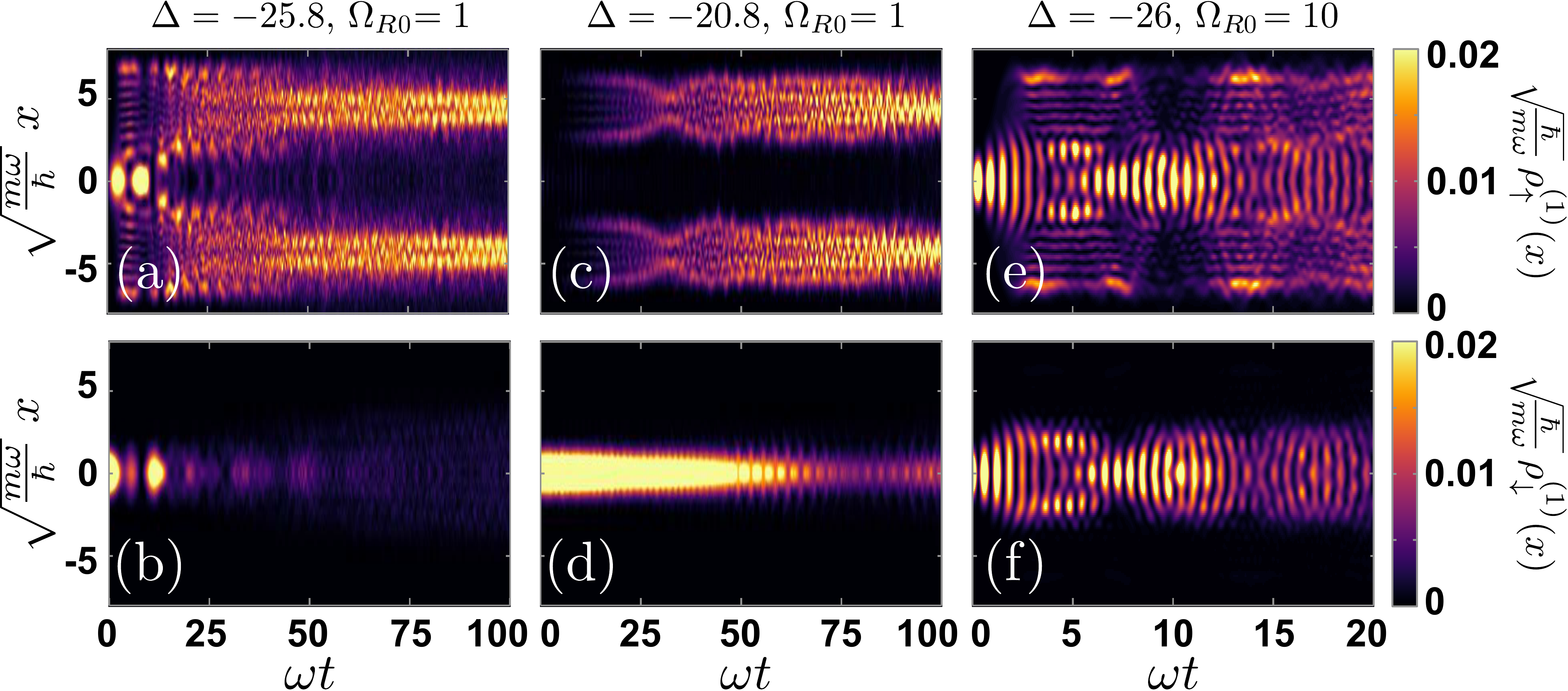}
\caption{Density evolution (a), (c), (e) $\rho^{(1)}_{\uparrow}(x;t)$ (spin-$\uparrow$) 
and (b), (d), (f) $\rho^{(1)}_{\downarrow}(x;t)$ (spin-$\downarrow$) of a single impurity for varying 
detuning $\Delta$ (see legends) 
and bare Rabi frequency (a)-(d) $\Omega_{R0}=1$ and (e), (f) $\Omega_{R0}=10$. 
It holds that $g_{B \downarrow}=0$ and $g_{B\uparrow}\equiv g_{BI}=1.5$, whilst the remaining parameters 
are the same to the ones considered 
in Fig.~\ref{fig:spectrum_weak_int}. 
The dynamics in (a)-(d) of the impurity spin-$\uparrow$ density manifests the TOC phenomenon accompanied by the depletion of the 
spin-$\downarrow$ state (see text). 
For $\Omega_{R0}=10$ the emergent Rabi oscillations are faster compared to the timescale where the impurity escapes the background, thus approaching the locally adiabatic spin-transfer regime. 
The considered pulse time corresponds to (a)-(d) $\tau=T=100>t$ and (e), (f) $\tau=T=20>t$. 
The Thomas-Fermi radius of the bosonic gas corresponds to $R_{TF}\approx 4.2$.}
\label{fig:density_strong_int_weak_dr} 
\end{figure*}

\begin{figure*}[ht]
\includegraphics[width=0.8\textwidth]{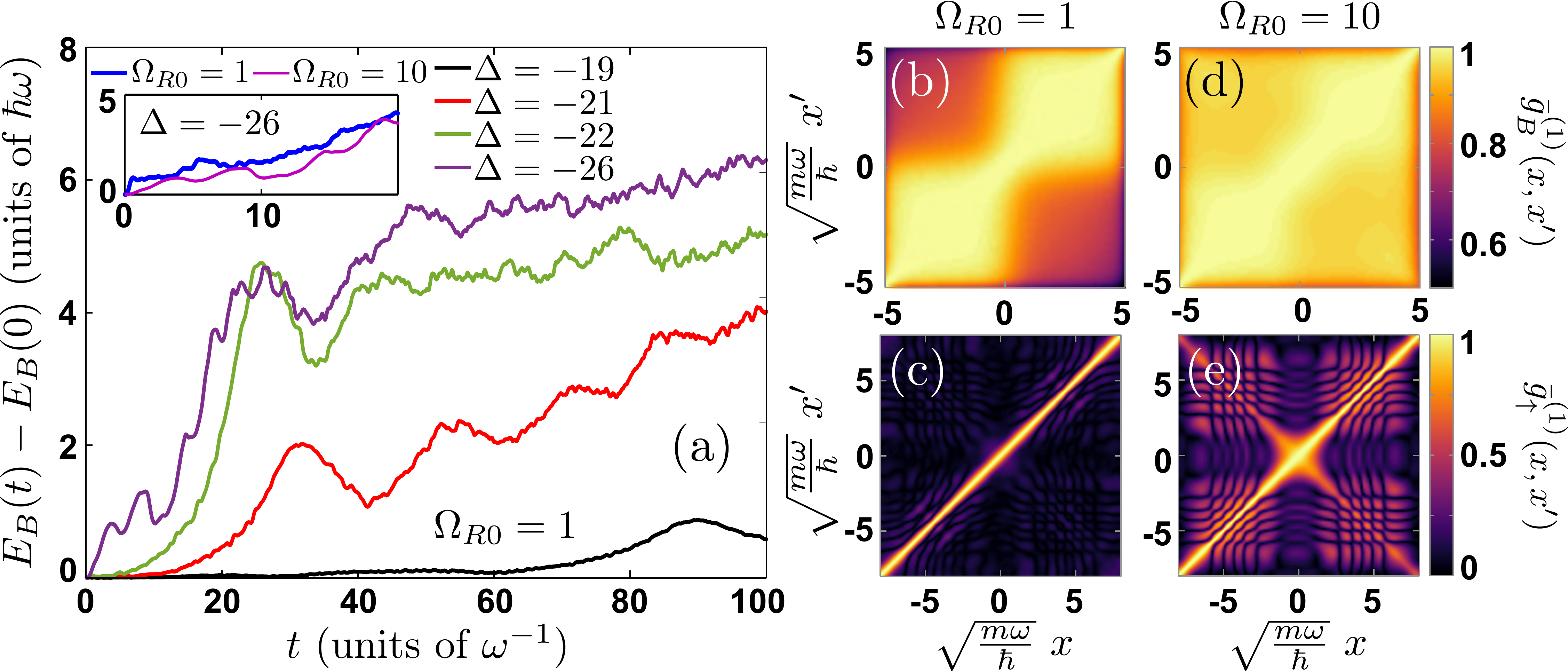}
\caption{Dynamics of the expectation value of the energy of the bosonic medium, $E_B(t)-E_B(0)$, for different 
detunings $\Delta$ (see legend), fixed Rabi frequency of the rf field $\Omega_{R0}=1$ and pulse time $\tau=T=100>t$. 
The observed increasing behavior with time showcases the energy transfer from the impurity to the BEC medium. 
The inset shows $E_B(t)-E_B(0)$ for constant $\Delta=-26$ and distinct values of $\Omega_{R0}$ (see legend). 
Time-averaged one-body coherence function of (b), (d) the bosonic medium and (c), (e) the impurity for varying 
$\Omega_{R0}$ (see legend) and 
fixed $\Delta=-26$ demonstrating coherence losses due to the TOC phenomenon.  
In all cases, $g_{B \downarrow}=0$, $g_{B\uparrow}\equiv g_{BI}=1.5$ and the remaining parameters are the same 
as in Fig.~\ref{fig:spectrum_weak_int}.}
\label{fig:energy_coherence_strong_int} 
\end{figure*}

The structure of $\bar{f}(\Delta)$ for $N_I=1$ can be understood by 
relying on the above-mentioned features of $V_{eff}(x;g_{BI})$. 
In particular, it is known~\cite{Mistakidis_orth_cat,Mistakidis_induced_cor} that the TOC occurs when the superfluidity of the BEC breaks due to rapidly moving 
impurities within its spatial extent that transfer their energy to the host by phonon emission. 
Accordingly, the spin-$\uparrow$ states created by an rf pulse with $\Delta>-25$ do not possess a large density overlap with 
the BEC and the TOC is not pronounced [see Fig.~\ref{fig:density_strong_int_weak_dr} (c) for $\Delta=-20.8$]. 
This fact permits their resolution as individual peaks in 
$\bar{f}(\Delta)$ for $N_I=1$ [Fig.~\ref{fig:spectrum_weak_int} (c) for $\Delta=-21$]. 
Notice that their larger amplitude when compared to the $V_{eff}(x;g_{BI})$ spectrum is attributed to 
their eventual but rather slow evolving TOC resulting in the population trapping of the impurity 
in the spin-$\uparrow$ state due to the transfer of the impurity energy to the background [Fig.~\ref{fig:energy_coherence_strong_int} (a)]. 
The situation is radically different for $\Delta \leq -25$ though, since in this case the transferred spin-$\uparrow$ 
impurities lie within the spatial extent of the BEC and are subsequently accelerated towards the periphery of the cloud due to 
$V_{eff}(x;g_{BI})$ [see Fig.~\ref{fig:density_strong_int_weak_dr} (a) for $\Delta=-25.8$ and $t<10$]. 
This process results in a pronounced and rapidly evolving TOC and leads to the accumulation of the spin-$\uparrow$ impurities 
outside of the Thomas-Fermi radius, implying $f(\Delta,\tau) \to 1$ for large evolution times ($t>25$) and consequently $\bar{f}(\Delta)$ 
is enhanced. 
Notice also that TOC is not spectrally selective as it appears as long as particles are transferred to the 
spin-$\uparrow$ state [Fig.~\ref{fig:spectrum_weak_int} (c) for $\tau>10$] and it is insensitive to the detuning from the resonance 
yielding a broad distribution instead of individual peaks [Fig.~\ref{fig:spectrum_weak_int_av} (c)]. 
This fact can be attributed to the rapid character of the TOC which occurs at a fast timescale 
of $\omega^{-1}$ as it has been recently revealed by pump-probe spectroscopy~\cite{Mistakidis_pump}. 

The additional spectral peaks identified in $\bar{f}(\Delta)$ for $\Delta < -26$ within 
$V_{eff}(x;g_{BI})$ [Fig.~\ref{fig:spectrum_weak_int_av} (c)] possess a secondary role. 
Indeed, the increased transfer rate to the spin-$\uparrow$ state associated with them results in a more efficient 
expulsion of the impurity from the spatial extent of the BEC [Fig.~\ref{fig:density_strong_int_weak_dr} (a)]. 
This process leads to a local maximum in $\bar{f}(\Delta)$ within the region of the continuum distribution of excitations. 
Turning our attention to $\bar{f}(\Delta)$ for $N_I=2$, we observe an almost perfect match with the spectrum of $N_I=1$. 
Small aberrations for $\Delta \leq -26$ can be attributed to a blockade-like phenomenon occurring during the dynamics 
which will be analyzed in Sec.~\ref{sec:two_body}. 
Let us note in passing that the kinetic energy of the impurities plays a crucial role in determining the spectral 
response of the system for $\Omega_{R0}=1$. 
As such, the diabatic approximation fails completely to capture the spectral features exhibited for such intensities of the rf pulse 
[Fig.~\ref{fig:spectrum_weak_int_av} (c)]. 

In order to visualize the impurity behavior we investigate the spatiotemporal evolution of its spin densities 
within the MB method, see Figs.~\ref{fig:density_strong_int_weak_dr} (a)-(d). 
Focusing on $\Delta_+^*=-25.8$ [Figs.~\ref{fig:density_strong_int_weak_dr} (a)-(b)] we observe the two transfer cycles 
of the impurity from its spin-$\downarrow$ to the spin-$\uparrow$ state and vice versa until $t\approx 12$. 
Afterwards the impurity remains trapped in its spin-$\uparrow$ state whilst the spin-$\downarrow$ one is almost 
completely depopulated. 
At the initial pulse times a Gaussian-like density hump builds upon $\rho_{\uparrow}^{(1)}(x;t)$ residing around 
$x=0$ which subsequently splits into two fragments that are symmetrically placed with respect to $x=0$ and support 
a multihump shape. 
Remarkably, these impurity density fragments execute a damped oscillatory motion during the pulse time and are 
located around the edges of the Thomas-Fermi radius of the BEC background, a process that gives rise to an 
impurity-medium phase-separation and accordingly leads to the TOC~\cite{Mistakidis_orth_cat,Mistakidis_induced_cor}. 
A similar phenomenology occurs also for other detuning corresponding to the remaining spectral resonances, 
see e.g. the time-evolution of the involved spin densities at $\Delta_+=-20.8$ depicted 
in Figs.~\ref{fig:density_strong_int_weak_dr} (c)-(d). 
Apparently, a much slower impurity transfer is achieved from the $\ket{\downarrow}$ to the 
$\ket{\uparrow}$ states with $\rho_{\downarrow}^{(1)}(x;t)$ lying at the trap center. 
However, also in this case, after the initial population of the $\ket{\uparrow}$ configuration 
by an impurity portion two density branches appear in $\rho_{\uparrow}^{(1)}(x;t)$ which undergo damped amplitude oscillations being centered at the edges of the Thomas-Fermi background. 
Notice here the humped structure of $\rho_{\uparrow}^{(1)}(x;t)$ which possesses a smaller number of 
nodes compared to $\Delta_+^*=-25.8$ indicating the lower excitation order of the impurity in the former case.  

The above-described TOC phenomenon is inherently related to a migration of the impurity energy into the BEC 
medium~\cite{Mistakidis_pump,Nielsen,Mukherjee_pulse}. 
This process can be directly verified by investigating the expectation value of the energy of the bosonic gas during 
the time-evolution, namely 
$E_B(t)\equiv \braket{\Psi(t)|\hat{H}_B^0+\hat{H}_{BB}|\Psi(t)}$, 
see also Eq.~(\ref{Htot_system}). 
The behavior of $E_B(t)-E_B(0)$ is presented in Fig.~\ref{fig:energy_coherence_strong_int} for $\Omega_{R0}=1$ 
and different detunings $\Delta$ of the rf field corresponding to specific resonant peaks identified in $\bar{f}(\Delta)$. 
Clearly the energy of the bath evinces an overall increasing tendency in the course of the pulse time implying an 
energy transfer from the impurity to its background. 
Of course, the respective impurity-BEC interaction energy i.e. 
$E_{BI}^{int}(t)\equiv \braket{\Psi(t)|\hat{H}_{BI}^{int}|\Psi(t)}$ simultaneously decreases (not shown here). 
The rate and amount of energy migration is more prominent for resonances characterized by a larger magnitude of 
detuning such that $\Delta_+\to \Delta_+^*\approx -26$ since it essentially yields a more significant spatial overlap of 
the involved metastable polaron state and the bath~\cite{Knap_catastr}. 
This increased overlap leads to a more rapid TOC and therefore transfer of the impurity 
energy to the bath. 
It is also worth mentioning that $E_B(t)-E_B(0)$ features certain amplitude oscillations which are attributed 
to the ``collisions'' of $\rho_{\uparrow}^{(1)}(x;t)$ with $\rho_{B}^{(1)}(x;t)$ during the phase-separation 
process~\cite{Mistakidis_orth_cat}. 
Moreover, keeping fixed the detuning e.g. to $\Delta_+^*=-26$ and considering a larger bare Rabi frequency leads 
to a smaller amount of energy transfer into the bath. 
This can be understood since the enhancement of $\Omega_{R0}$, and thus of the pulse amplitude, essentially gives rise to a faster spin-transfer which 
naturally does not let the impurity to adapt to modified coupling with the environment~\cite{Schmidt_rev,Liu_rf1}. 

To appreciate the reduced degree of coherence already indicated by the rf spectrum we rely on the time-average 
over the entire evolution (T), see also the discussion below Eq.~(\ref{eq:single_par_den_matr}), of the spatial 
first-order coherence function
\begin{equation}
\bar{g}^{(1)}_{\sigma}(x,x')=\frac{1}{T}\int_0^T dt \frac{\rho^{(1)}_{\sigma}(x,x';t)}{\sqrt{\rho^{(1)}_{\sigma}(x;t) 
\rho^{(1)}_{\sigma}(x';t)}}.\label{coherence_average} 
\end{equation}
In this expression, $\rho^{(1)}_{\sigma}(x,x';t)$ refers to the $\sigma=B,\uparrow,\downarrow$ component 
one-body reduced density matrix [Eq.~(\ref{eq:single_par_den_matr})] whose diagonal is the one-body density 
i.e. $\rho^{(1)}_{\sigma}(x;t)$. 
Evidently, $|\bar{g}^{(1)}_{\sigma}(x,x';t)| \in [0, 1]$ reveals the spatially resolved deviation of a MB wave function 
from a corresponding product state. 
In particular, when $|\bar{g}^{(1)}_{\sigma}(x,x')|=1$ the component is called fully coherent otherwise coherence 
losses come into play exposing the build up of correlations~\cite{mistakidis_phase_sep}. 
We remark that initially ($t=0$) both the impurity and the environment are almost fully coherent (not shown for brevity).  
Inspecting $|\bar{g}^{(1)}_{B}(x,x')|$ in the case of $\Omega_{R0}=1$ dictates that in the bosonic gas a small 
amount of coherence losses occur as can be inferred from the corresponding off-diagonal since e.g. 
$|\bar{g}^{(1)}_{B}(x=-4,x'=4)|\approx 0.85$ [Fig.~\ref{fig:energy_coherence_strong_int} (b)]. 
However, regarding the impurity we can readily deduce the occurrence of substantial coherence losses being 
evident by the fade out of $|\bar{g}^{(1)}_{\uparrow}(x,x'\neq x)|\to 0$ [Fig.~\ref{fig:energy_coherence_strong_int} (c)] 
and therefore any polaronic notion is lost as also probed by the rf spectra [Fig.~\ref{fig:spectrum_weak_int} (c)]. 
A similar conclusion has also been unveiled recently by monitoring the polaronic spectrum using a 
pump-probe spectroscopic scheme~\cite{Mistakidis_pump}. 

Increasing the bare Rabi frequency, namely $\omega \ll  \Omega_{R0}=10 < c/ \xi $, results in a modified spectral response 
$f(\Delta;\tau)$ [Fig.~\ref{fig:spectrum_weak_int} (d)] and therefore time-averaged rf spectrum $\bar{f}(\Delta)$ 
[Fig.~\ref{fig:spectrum_weak_int_av} (d)]. 
In the course of the pulse time $f(\Delta;\tau)$ initially ($t<4$) performs single frequency Rabi oscillations and 
later on features a multifrequency behavior [Fig.~\ref{fig:spectrum_weak_int} (d)]. 
Several resonant peaks e.g. at $\Delta_+\approx-26$, $\Delta_+\approx-20$, $\Delta_+\approx-17.3$ and $\Delta_+\approx-14.5$ 
appear in both $f(\Delta;\tau)$ and $\bar{f}(\Delta)$ designating motionally excited polaronic states formed by the partially dressed 
portion of the spin-$\uparrow$ impurity. 
This partial dressing mechanism of the impurity can be directly inferred from the spatially resolved evolution of the 
spin densities, see for instance Figs.~\ref{fig:density_strong_int_weak_dr} (e), (f). 
Indeed, at the initial stages of the dynamics ($t<2.5$) a fast Rabi-type transfer of atoms from the spin-$\uparrow$ to 
the spin-$\downarrow$ 
state and vice versa occurs causing an undulation to the BEC density in the vicinity of $\rho_{\uparrow}^{(1)}(x;t)$ within  
$\Omega_{R0}^{-1}> \xi /c$. 
Notice that in the short time-interval where the impurity is excited to the spin-$\uparrow$ state ($\sim \pi/ \Omega_{R0} \approx 0.3<\omega$) 
it can hardly disperse until it is de-excited to the spin-$\downarrow$ state. 
Due to this non-dispersive character of $\rho_{\uparrow}^{(1)}(x;t)$ in each Rabi-cycle the impurity interacts stronger 
with the emitted BEC excitations (phonons) leading to its larger spatial localization in the background when compared to 
the case of $\Omega_{R0}=1$ [Fig.~\ref{fig:density_strong_int_weak_dr} (a), (b)]. 
At longer evolution times a portion of $\rho_{\uparrow}^{(1)}(x;t)$ can escape to the periphery of the BEC leading to a gradual delocalization of $\rho_{\uparrow}^{(1)}(x;t)$ which later on contracts towards the trap 
center and conducts afterwards a similar behavior. 
Due to the impurity migration among the spin states $\rho_{\downarrow}^{(1)}(x;t)$ has a complementary structure to 
$\rho_{\uparrow}^{(1)}(x;t)$ developing humps in the location of the $\rho_{\uparrow}^{(1)}(x;t)$ dips while
remaining within the spatial extension of the medium. 
The above-effects crucially depend on the timescale of the development of impurity-BEC correlations ($\sim \xi/c$) when 
compared to the corresponding one set by the trap ($\sim \omega^{-1}$), thus exposing the emergence of the locally 
adiabatic regime. 
Moreover, the medium is almost perfectly coherent since $|\bar{g}^{(1)}_{B}(x,x'\neq x)|\approx 0.95$ 
[Fig.~\ref{fig:energy_coherence_strong_int} (d)]. 
However, the spin-$\uparrow$ impurity experiences more severe coherence losses [Fig.~\ref{fig:energy_coherence_strong_int} (e)]. 
Indeed, the impurity portion remaining within the bosonic background is adequately coherent 
namely $|\bar{g}^{(1)}_{\uparrow}(-1.7<x<1.7,-1.7<x'<1.7)| \approx 0.8$ while the segments escaping the medium are less 
coherent within themselves and between each other, see for instance 
$|\bar{g}^{(1)}_{\uparrow}(2<x<7,2<x'<7)| \approx 0.5$. 
Note also that the impurity portion escaping the medium is almost fully incoherent with the one remaining inside its spatial extent.

By comparing the time-averaged spectra in  [Fig.~\ref{fig:spectrum_weak_int_av} (d)], 
we observe that the $V_{eff}(x;g_{BI})$ method is able to predict the location of the polaronic resonances emerging 
for $\Delta>-26$ while it neglects those resonances appearing for $\Delta<-26$ e.g. at $\Delta_+\approx-28.4$. 
Also, $\bar{f}(\Delta)$ obtained using $V_{eff}(x;g_{BI})$ has an overall lower amplitude when compared to the MB case.  
This is attributed to the incoherent fraction of impurity atoms exhibiting TOC, a process that is not 
captured within $V_{eff}(x;g_{BI})$, similarly to the $\Omega_{R0}=1$ case.  
The resonant peaks building upon $\bar{f}(\Delta)$ in the MB scenario are present also in the case of two non-interacting 
bosonic impurities with the ones appearing for $\Delta>-27$ being slightly shifted and those for $\Delta<-26$ having a 
relatively larger amplitude than for $N_I=1$. 
Here, the existence of impurity-impurity induced interactions materializes in a shift of the spectral lines whilst 
their higher amplitude is most probably caused by the smaller lifetime of the impurity excitations when $N_I=2$ due to a more pronounced TOC. 
We note that the multi-peaked structure of $\bar{f}(\Delta)$ witnesses that $\Omega_{R0}=10$ here is proximal but  can not be classified 
into the diabatic spin-transfer regime since it substantially deviates from a Lorentzian distribution. 
Recall that this observation is in sharp contrast to the $g_{BI}=0.5$ scenario [Fig.~\ref{fig:spectrum_weak_int} (e)] as well as 
to $g_{BI}=-0.5$ that will be shown later on [Fig.~\ref{fig:spectrum_weak_int_av} (f)] where in both cases $\bar{f}(\Delta)$ at $\Omega_{R0}=10$ 
represents a Lorentzian being in good agreement with the outcome of the diabatic approximation. 
As a consequence, we can deduce that for increasing impurity-medium interaction strength the diabatic spin-transfer regime is 
entered for larger bare Rabi frequencies. 
This result is attributed to the prominent role of impurity-phonon correlations developing 
for $t \approx \xi /c$. 

The diabatic spin-transfer regime is indeed realized for increasing Rabi frequency $\Omega_{R0}\gg c/ \xi$. 
A paradigmatic example is provided here e.g. for $\Omega_{R0}=40$ where as it can be seen in the inset of 
Fig.~\ref{fig:spectrum_weak_int_av} (d) $\bar{f}(\Delta)$ within the MB approach has a Lorentzian shape for 
both $N_I=1$ and $N_I=2$ and it coincides with the prediction of the diabatic approximation. 
Apparently, $V_{eff}(x;g_{BI})$ yields the same outcome, suggesting that MB excitation channels are 
strongly suppressed. 
The polaronic resonance in all cases is positioned at $\Delta_+^*\approx -24$.

To illustrate how the spectrum is modified with the pulse intensity, we depict $\bar{f}(\Delta)$ 
for varying $\Omega_{R0}$ in Fig.~\ref{fig:spectrum_weak_int_av_dif_freq} (b). 
As expected, a power broadening of $\bar{f}(\Delta)$ takes place for increasing $\Omega_{R0}$ leading to a perfectly 
shaped Lorentzian distribution for $\Omega_{R0}\gg c/ \xi$, see the inset of Fig.~\ref{fig:spectrum_weak_int_av_dif_freq} (b). 
The latter is a case example of the diabatic spin-transfer regime [inset of Fig.~\ref{fig:spectrum_weak_int_av} (d)]. 
In contrast, for small $\Omega_{R0}$, satisfying $\Omega_{R0}\leq c/ \xi$, the spectral line $\bar{f}(\Delta)$ has several 
peaks which correspond to distinct polaronic states. 
Accordingly, these resonances become narrower and their position is shifted [hardly visible in Fig.~\ref{fig:spectrum_weak_int_av_dif_freq} (b)] 
for decreasing $\Omega_{R0}$, e.g. see $\bar{f}(\Delta)$ for $\Omega_{R0}=0.1$ and $\Omega_{R0}=10$.

\subsection{Attractive Bose polaron: mode-coupling and signatures 
of induced interactions }\label{sec:attract_int}

\begin{figure*}[ht]
\includegraphics[width=0.8\textwidth]{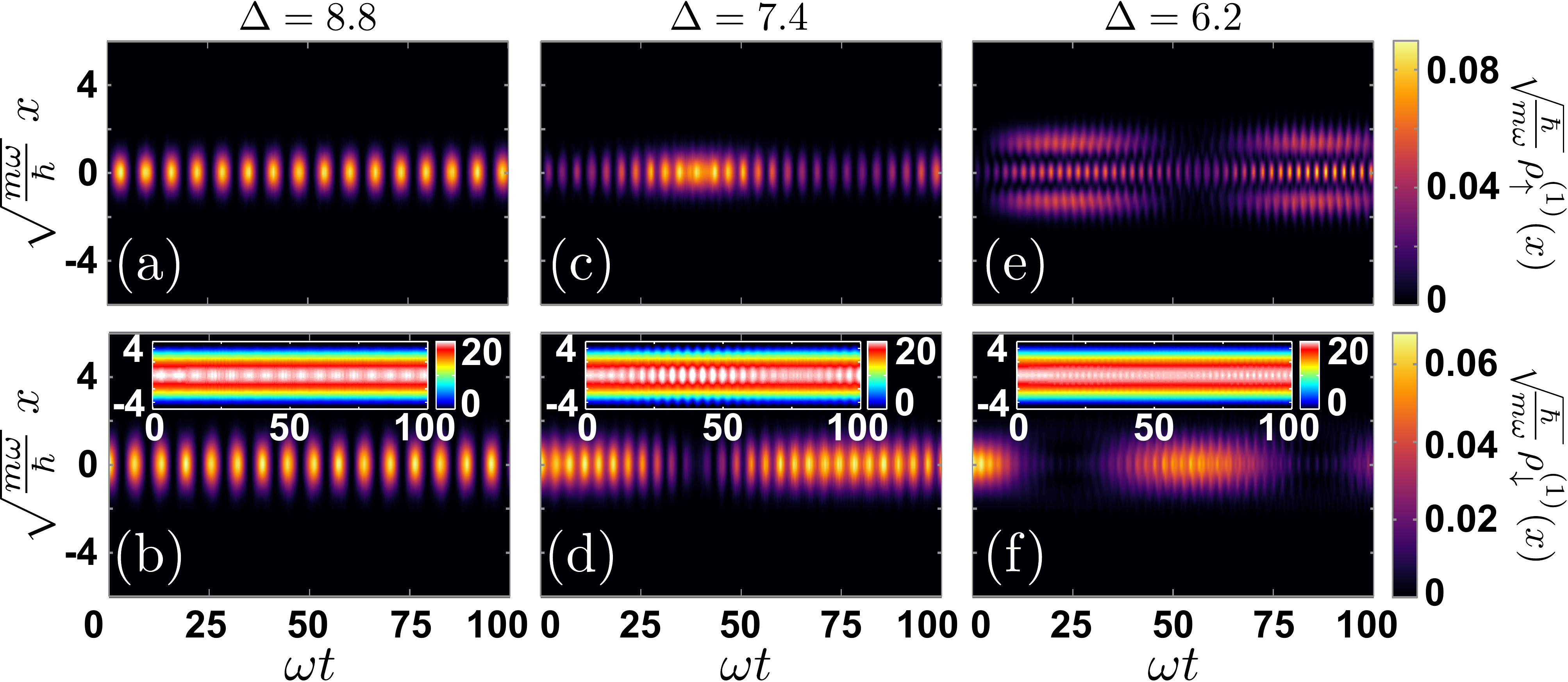}
\caption{Spatiotemporal evolution of the density (a), (c), (e) $\rho^{(1)}_{\uparrow}(x;t)$ and (b), (d), (f) 
$\rho^{(1)}_{\downarrow}(x;t)$ of a single impurity for distinct detunings $\Delta$ of the rf field (see legends) and 
constant bare Rabi frequency $\Omega_{R0}=1$ visualizing different motionally excited states of the attractive polaron. 
The insets in (b), (d), (f) show the corresponding density evolution $\rho^{(1)}_{B}(x;t)$ of the bosonic medium featuring 
(b), (f) phonon emission and (d) an excited breathing mode.  
The interaction strengths of the spinor parts of the impurity with the background are $g_{B \downarrow}=0$ 
and $g_{B\uparrow}\equiv g_{BI}=-0.5$. 
Other system parameters are considered to be the same as in Fig.~\ref{fig:spectrum_weak_int}. 
In all cases the employed pulse time refers to $\tau=T=100>t$, while the Thomas-Fermi radius of the background is 
$R_{TF}\approx 4.2$.}
\label{fig:density_neg_int} 
\end{figure*}

Next we aim to study attractive impurity-medium coupling, e. g. $g_{BI}=-0.5$. 
The resulting spectral response $f(\Delta;\tau)$ and its time-average $\bar{f}(\Delta)$ for $\Omega_{R0}=\omega=1 \ll c/ \xi$ are 
provided in Fig.~\ref{fig:spectrum_weak_int} (e) and Fig.~\ref{fig:spectrum_weak_int_av} (e) respectively. 
A polaronic resonance located at $\Delta_+^*=8.8$ appears in the rf signal, characterized by a single Rabi frequency $\Omega_R (\Delta_+^*) \approx 0.99$, see the inset in Fig.~\ref{fig:spectrum_weak_int} (e) where the Fourier 
spectrum of $f(\Delta;\tau)$ is depicted. 
This resonance is the energetically lowest interacting state of the spin-$\uparrow$ impurity with the bath and manifests as 
the formation of an attractive Bose polaron~\cite{Massignan,Mistakidis_induced_cor}. 
The dressed Rabi frequency is  $\Omega_R(\Delta)=\sqrt{(Z\Omega_{R0})^2+(\Delta-\Delta_+^*)^2}$ with 
$Z\approx 0.99$, as identified in the spectrum. 
The underlying polaron generation mechanism can also be inferred from the impurity spin densities which undergo single 
frequency Rabi oscillations among each other [Figs.~\ref{fig:density_neg_int} (a), (b)]. 
Indeed, the periodic revival of density lumps in $\rho_{\uparrow}^{(1)}(x;t)$ inside the medium signals the polaron 
formation which subsequently causes small amplitude perturbations in $\rho_{B}^{(1)}(x;t)$ 
[inset of Fig.~\ref{fig:density_neg_int} (b)] due to weakly excited phononic modes triggered by the back-action of the impurity to its environment, see also Ref.~\cite{Mistakidis_inject_imp}.

Additional peaks at smaller detuning in $f(\Delta;\tau)$, refer either to higher-lying polaronic states or to resonant coupling between the impurity motion and the collective excitations of the BEC environment~\cite{Dolgirev}. 
It is worth mentioning that the non-Lorentzian shape of the
quasi-particle resonances is an indication that  $\Omega_{R0}=1$ Rabi frequency does not lead to a diabatic spin-transfer; compare the spectral line for the $N_I=1$ MB case and the diabatic approximation shown 
in Fig.~\ref{fig:spectrum_weak_int_av} (e). 

To explicitly identify the polaronic resonances, we utilize the effective potential model which here for the 
spin-$\uparrow$ impurity corresponds to a harmonic oscillator of effective frequency $\omega_{eff}=\sqrt{1+(\abs{g_{BI}}/g_{BB})}\approx \sqrt{2} \omega$ within the Thomas-Fermi approximation \cite{Mistakidis_induced_cor}. 
The lowest-lying state have energies  $\epsilon_n\approx \hbar\sqrt{2} \omega (n+1/2)$, where 
the index $n$ denotes each eigenstate. The respective energy differences are of the order of $\delta \epsilon \approx \sqrt{2} \gtrapprox \Omega_{R0}=1$, meaning that the applied pulse enables the resolution of the individual eigenstates of $V_{eff}(x;g_{BI})$. 

Employing the time-averaged spectrum $\bar{f}(\Delta)$ within $V_{eff}(x;g_{BI})$ [Fig.~\ref{fig:spectrum_weak_int_av} (e)] 
we indeed verify the existence of the individual polaronic states whose positions are in  perfect agreement with those 
predicted by the MB approach. 
Particularly, the peaks at $\Delta_+\approx 6.2$, $\Delta_+\approx 5.4$ and $\Delta_+\approx 3.2$ 
signify motionally excited states of the polaron and display a Rabi frequency 
$\Omega_R(\Delta)$ which is smaller to the one for $\Delta_+^*=8.8$ [see the dashed box in the inset of Fig.~\ref{fig:spectrum_weak_int} (e)]. 
This decreasing behavior of $\Omega_R$ is attributed to the reduced overlap of these states, and hence reduced quasi-particle residue, with the initial non-interacting impurity-bath MB state~\cite{Massignan}.  
The motionally excited nature of such states can be readily deduced e.g. from the three-peaked structure of the spin-$\uparrow$ 
impurity density during the pulse time, see Fig.~\ref{fig:density_neg_int} (e). 
At such detuning, a slow population transfer from the spin-$\downarrow$ to the spin-$\uparrow$ state and 
vice versa occurs with the $\rho_{\downarrow}^{(1)}(x;t)$ residing around the trap center 
[Fig.~\ref{fig:density_neg_int} (e), (f)]. 
Also, the motion of the spin-$\uparrow$ impurity portion weakly disturbs $\rho_{B}^{(1)}(x;t)$, which subsequently emits small amplitude sound waves [inset of Fig.~\ref{fig:density_neg_int} (f)].  

On the other hand, the spectral peaks e.g. at $\Delta_+\approx 7.4$ in $\bar{f}(\Delta)$ that are not captured 
with $V_{eff}(x;g_{BI})$, are resonances for the impurity motion with the collective excitations of the BEC medium~\cite{Dolgirev,Mistakidis_induced_cor}, see Fig.~\ref{fig:spectrum_weak_int_av} (e). 
This mode-coupling can be demonstrated, for instance, by monitoring the underlying densities 
of the individual components [Figs.~\ref{fig:density_neg_int} (c), (d)]. 
A particle transfer between the $\ket{\uparrow}$ and the $\ket{\downarrow}$ impurity states is evident, accompanied by a breathing motion of $\rho_{\uparrow}^{(1)}(x;t)$ and $\rho_{\downarrow}^{(1)}(x;t)$ which 
lie near the trap center. 
Most importantly, when the population of the polaron $\ket{\uparrow}$ state is dominant, $\rho_{B}^{(1)}(x;t)$ 
exhibits an enhanced breathing dynamics due to the impurity back-action [inset of Fig.~\ref{fig:density_neg_int} (d)].

Turning to the spectral response of two non-interacting bosonic impurities in the MB case we observe that the structural 
configuration of $\bar{f}(\Delta)$ is almost the same to the one for $N_I=1$, see Fig.~\ref{fig:spectrum_weak_int_av} (e). 
This indicates that the majority of the polaronic resonances is of single-particle origin. 
However, by closely inspecting $\bar{f}(\Delta)$ it becomes evident that the position of some of the emergent peaks 
referring to polaronic excitations, e.g. located at $\Delta_+=5.8$, $\Delta_+=4.2$ are shifted towards larger 
detunings when compared to the $N_I=1$ scenario by an average amount of $\delta \Delta_+ \approx 0.4$. 
This latter behavior suggests the appearance of attractive induced impurity-impurity interactions mediated by 
the bosonic gas~\cite{Mora,Mistakidis_Volosniev_induce_int,induced_int_artem,Scazza}. 
Similar manifestations of induced interactions have been revealed in the spectrum of the contrast of bosonic 
impurities obtained via the Ramsey response~\cite{Mistakidis_induced_cor}. 
Let us note is passing that the two-body character of the polaronic resonances will be further discussed in the 
next section~\ref{sec:two_body}. 

The rf spectrum for a larger bare Rabi frequency such that $\omega \ll \Omega_{R0}=10< c/ \xi$ but still $g_{BI}=-0.5$ 
changes drastically when compared to the $\Omega_{R0}\leq \omega$ scenario. 
Monitoring $f(\Delta;\tau)$ we observe the appearance of a single polaronic resonance located at $\Delta_+^*\approx 8.8$ 
which Rabi oscillates in the course of the pulse time [Fig.~\ref{fig:spectrum_weak_int} (f)] while possessing 
a frequency that obeys $\Omega_R(\Delta)=\sqrt{(Z\Omega_{R0})^2+(\Delta-\Delta_+^*)^2}$ with $Z\approx 1$ as shown in the inset of Fig.~\ref{fig:spectrum_weak_int} (b). 
The time-averaged signal $\bar{f}(\Delta)$ is Lorentzian and coincides between $N_I=1$ and $N_I=2$. 
In this sense, two-body processes are here suppressed. 
Also, the effective potential approximation for $N_I=1$ provides exactly the same $\bar{f}(\Delta)$ indicating that 
MB processes are negligible. 
To exemplify whether the diabatic spin-transfer regime has been realized we compare $\bar{f}(\Delta)$ among the MB method and 
the diabatic approximation [Fig.~\ref{fig:spectrum_weak_int_av} (f)]. 
The two spectra are almost equivalent and thus $\Omega_{R0}=10< c/ \xi$ essentially acts as a diabatic pulse, a result that stems from the weakly attractive impurity-BEC interactions. 

Inspecting $\bar{f}(\Delta)$ for various $\Omega_{R0}$ it allows us to infer the adiabatic and the diabatic 
nature of the employed 
rf pulse [Fig.~\ref{fig:spectrum_weak_int_av_dif_freq} (c)]. 
Indeed, for $\Omega_{R0}\gg \omega$ the time-averaged spectral line as captured by $\bar{f}(\Delta)$ has a  
Lorentzian form indicating its proximity to the intense pulse case. 
In addition, it exhibits a single polaronic peak whose position depends on $\Omega_{R0}$ and as a result of the power broadening 
its width increases for a larger $\Omega_{R0}$. 
However, in the case of adiabatic spin-transfer ($\Omega_{R0}\leq \omega$) the spectrum is substantially overall narrower 
and most importantly 
it shows a multitude of polaronic resonances for $\Delta<\Delta_+^*$ corresponding to energetically higher-lying excitations of these 
quasi-particles. 
For instance, as discussed above considering $\Omega_{R0}=1$ a variety of resonances appear in $\bar{f}(\Delta)$ while for $\Omega_{R0}=0.1$ only a single narrow peak occurs. 
The latter observation can be attributed to the fact that the number of the resonant peaks when $\Omega_{R0}\leq \omega$ 
is of course 
reduced for smaller $\Omega_{R0}$ since then the intensity of the pulse decreases and thus it is less probable to 
populate higher-lying motionally excited states of the impurity MB spectrum.

\section{Characteristic two-body polaron processes: Rabi oscillations and correlation induced dephasing}\label{sec:two_body}

To elucidate the existence of possible two-body correlation mechanisms that contribute in the polaron dynamics (in the presence of two impurities) for 
varying Rabi frequency and detuning of the rf field we invoke the spin-spin probability 
\begin{equation}
P_{aa'}(t)=\braket{\Psi(t)|\hat{P}_{aa'}|\Psi(t)}.\label{two_body_prob}
\end{equation}
The participating spin operators read $\hat{P}_{\uparrow \uparrow}=\ket{1,1}\bra{1,1}$, 
$\hat{P}_{\downarrow \downarrow}=\ket{1,-1}\bra{1,-1}$ and $\hat{P}_{\uparrow \downarrow}=\ket{1,0}\bra{1,0}+\ket{0,0}\bra{0,0}$, 
where the spin basis $\ket{S,S_z}$ is used. 
These probabilities satisfy $P_{\uparrow \uparrow}(t)+P_{\downarrow \downarrow}(t)+P_{\uparrow \downarrow}(t)=1$. 
Apparently, the limiting cases $P_{\uparrow \uparrow}(t)=1$ and $P_{\uparrow \downarrow}(t)=1$ refer to the 
formation of a two and a single polaron state respectively, whilst $P_{\downarrow \downarrow}(t)=1$ signifies 
the absence of polaronic excitations. 
Below we discuss the behavior of $P_{aa'}(t)$ exemplarily for specific impurity-medium interaction strengths, 
bare Rabi frequencies (corresponding to adiabatic or diabatic spin-transfer) and different detunings of the rf scheme. 
Especially by focusing on parameter regions ($\Omega_{R0}$, $\Delta$) where the spectral line peaks 
between $N_I=1$ and $N_I=2$ impurities do not differ we showcase that even in these cases two-body mechanisms 
play a role for the adequate interpretation of the polaron dynamics.  
\begin{figure*}[ht]
\includegraphics[width=0.9\textwidth]{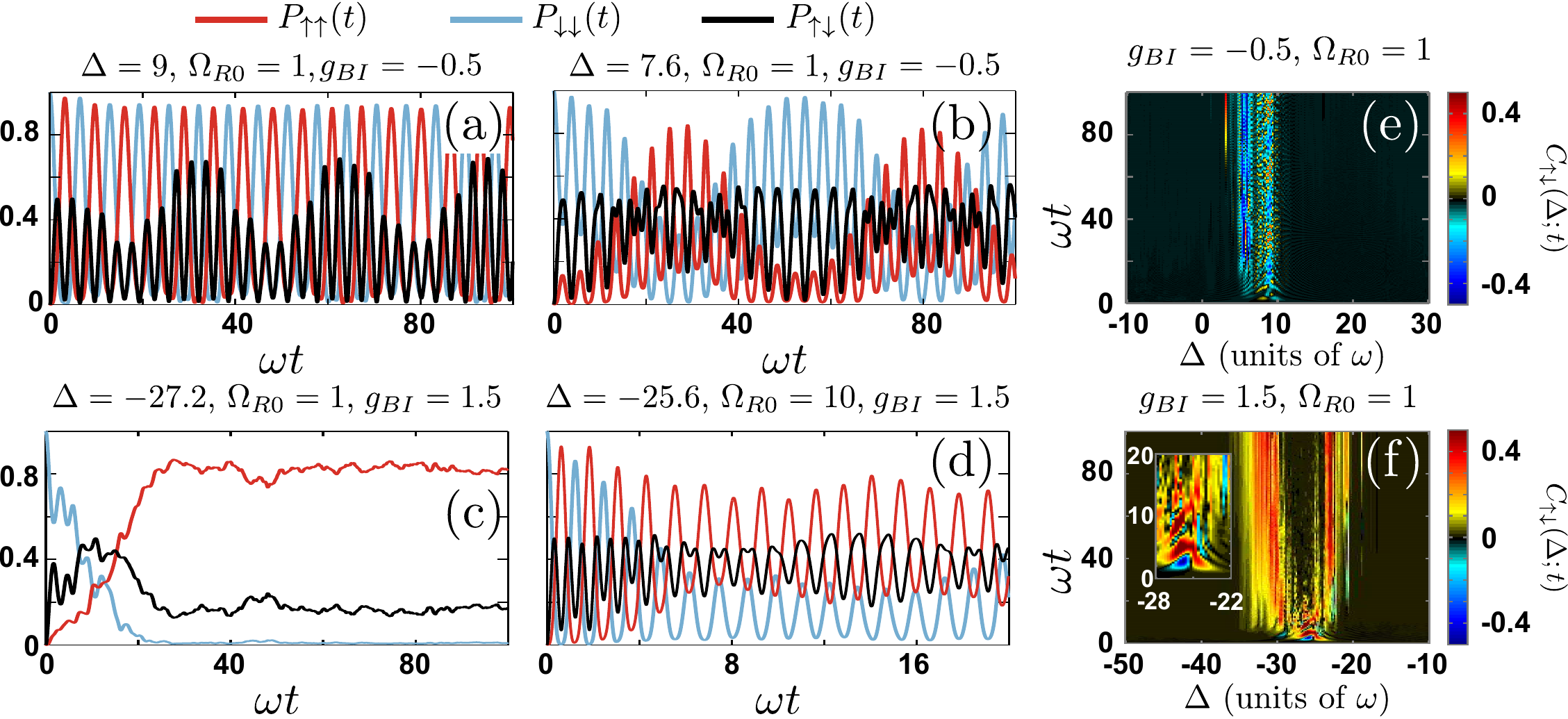}
\caption{(a)-(d) Dynamics of the two-body probability [Eq.~(\ref{two_body_prob})] for the two non-interacting bosonic 
impurities to reside both in the spin-$\uparrow$ or the 
spin-$\downarrow$ state or one in the spin-$\uparrow$ and the other in their spin-$\downarrow$ state for 
selective detunings and 
bare Rabi frequencies of the rf field as well as impurity-medium interaction strengths (see legends). 
The non-constant amplitude of $P_{\uparrow \downarrow}(t)$ signals the development of polaron-polaron induced correlations. 
Time-evolution of the spin correlator [Eq.~(\ref{spin_corel})] for (e) $g_{BI}=-0.5$, (f) $g_{BI}=1.5$ with 
respect to the detuning $\Delta$ of the rf field characterizing the correlated spin dynamics of the impurities.  
The inset of (f) shows the short-time dynamics of the corresponding spin correlator. 
The two impurities are non-interacting and the bath is composed of $N_B=100$ bosons with $g_{BB}=0.5$. 
The multicomponent setting is mass-balanced, $m_B=m_I$, and it is confined in a harmonic 
trap of frequency $\omega=1$. 
The pulse time in (a), (b), (c), (e), (f) $\tau=T=100>t$ while in (d) $\tau=T=20>t$.}
\label{fig:two_body_prob} 
\end{figure*} 

Regarding attractive impurity-bath couplings namely $g_{BI}=-0.5$ we present $P_{aa'}(t)$ for $\Omega_{R0}=1$ and 
$\Delta=9$ in Fig.~\ref{fig:two_body_prob} (a). 
Recall that this parameter choice corresponds to the lowest-lying polaron state in $f(\Delta;\tau)$ while the 
resonance position remains the same for $N_I=1$ and $N_I=2$, see also Fig.~\ref{fig:spectrum_weak_int} (e). 
An almost perfect precession dynamics from the $\ket{\downarrow \downarrow}$ [$P_{\downarrow \downarrow}(t)$] 
to the $\ket{\uparrow \uparrow}$ [$P_{\uparrow \uparrow}(t)$] spin states and vice versa takes place, a process 
that gives rise to the periodic formation of a two polaron state. 
Notice the small reduction of the amplitude of both $P_{\downarrow \downarrow}(t)$ and $P_{\uparrow \uparrow}(t)$ 
in the course of time which is indicative of a correlation induced dephasing mechanism. 
Simultaneously, $P_{\uparrow \downarrow}(t)$ oscillates with an almost $\pi/4$ phase difference with respect to both $P_{\downarrow \downarrow}(t)$ and $P_{\uparrow \uparrow}(t)$ but most importantly it exhibits a 
modulated amplitude, i.e. a beating behavior. 
The latter observation hints towards the presence of weak impurity-impurity correlations mediated by the bath 
since in the absence of impurity correlations $P_{\uparrow \downarrow}(t)$ would perform constant amplitude 
oscillations according to 
$P_{\uparrow \downarrow} (t)=2 A(t) [1-A(t)] \sin^2(Z \Omega_{R0} t/2 ) + (1/2) A^2(t) \sin^2 (Z \Omega_{R0} t)$ 
with $A(t)$ denoting the amplitude of the $f(\Delta;\tau)$ oscillations~\cite{Scully}.  
\begin{figure*}[ht]
\includegraphics[width=0.8\textwidth]{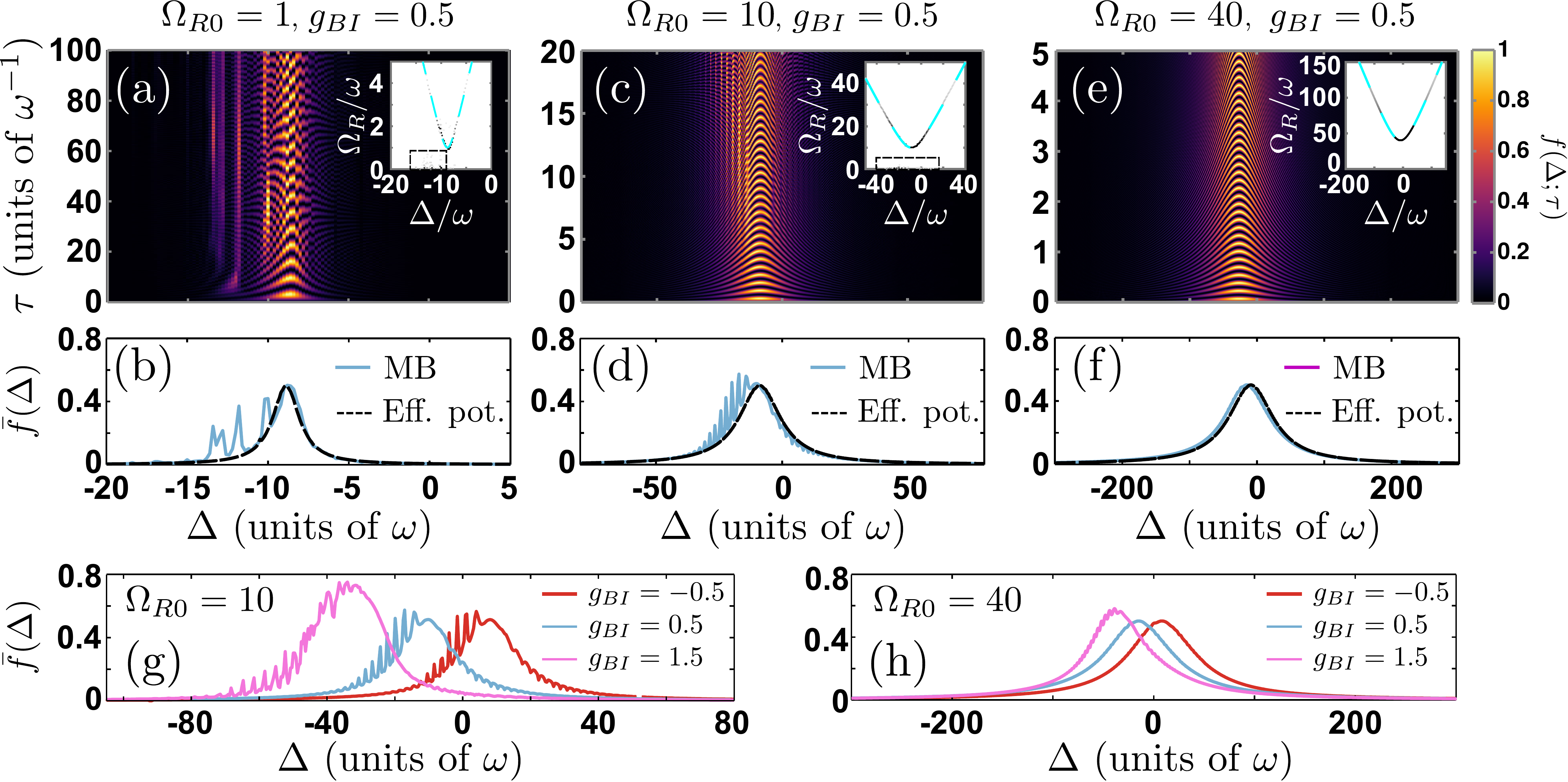}
\caption{Time-evolution of the spectroscopic signal $f(\Delta;\tau)$ of a heavy single impurity $N_I=1$ for 
varying detuning $\Delta$ of 
the rf field and repulsive impurity-bath interactions $g_{BI}=0.5$ using a bare Rabi frequency (a) $\Omega_{R0}=1$, 
(c) $\Omega_{R0}=10$ 
and (e) $\Omega_{R0}=40$ quantifying the crossover between the distinct spin-transfer regimes. 
A side peak structure manifests in $f(\Delta;\tau)$ due to the mode-coupling between the polarons and 
collective excitations of the bath (see text). 
Insets showcase the spectrum $f(\Delta;\Omega_R)$ whilst the dashed 
lines provide $\Omega_R(\Delta)=\sqrt{(Z\Omega_{R0})^2+(\Delta-\Delta_+^*)^2}$ with  
$\Delta_+^*\approx -8.8$ referring to the position of the resonance and $Z\approx 0.98$ being the quasi-particle residue. 
Dashed boxes indicate the resolved frequencies $\Omega_R$ of the spectral peaks at $\Delta<\Delta_+^*$. 
Time-averaged spectroscopic signal $\bar{f}(\Delta)$ within the MB method and the effective potential 
approximation (see legends) for 
(b) $\Omega_{R0}=1$, (d) $\Omega_{R0}=10$ and (f) $\Omega_{R0}=40$. 
$\bar{f}(\Delta)$ obtained in the MB approach for different impurity-medium interaction strengths 
(see legend) for (g) $\Omega_{R0}=10$ and 
(h) $\Omega_{R0}=40$. 
The system is mass-imbalanced, i.e. $m_I=(174/7)m_B$, and confined in a harmonic trap of $\omega=1$. 
It consists of $N_B=100$ bosons with $g_{BB}=0.5$ and a single spinor impurity with $g_{B\downarrow}=0$ and 
$g_{B\uparrow}\equiv g_{BI}$ indicated in the respective legends.}
\label{fig:spectrum_heavy_imp} 
\end{figure*} 

Keeping fixed $\Omega_{R0}=1$ and $g_{BI}=-0.5$ but moving to the spectral peak located at $\Delta_+=7.6$ which 
alludes to the resonance of the impurity motion with the background breathing mode we find that the ensuing 
dynamics of $P_{aa'}(t)$ is much more complex than for $\Delta=9=\Delta_+^*$ [Fig.~\ref{fig:two_body_prob} (b)] and it 
contains two predominant frequencies. 
Indeed, the response of $P_{\downarrow \downarrow}(t)$ and $P_{\uparrow \uparrow}(t)$ evinces a slow drift 
from the $\ket{\downarrow \downarrow}$ to the $\ket{\uparrow \uparrow}$ state with a period $T_1\approx 55$. 
The timescale of this process is in line with the one referring to the maximal occupation of $\rho_{\uparrow}^{(1)}(x;t)$, 
see e.g. Figs.~\ref{fig:density_neg_int} (c), (d) when $N_I=1$ 
and note that for $N_I=2$ the relevant time-interval exhibits a slight shift due to the impurity-impurity effective interactions.  
The probabilities $P_{aa'}(t)$ exhibit additional faster oscillations having a significantly larger frequency and 
testifying a population transfer from $P_{\downarrow \downarrow}(t)$ towards $P_{\uparrow \downarrow}(t)$ 
(single-polaron) and $P_{\uparrow \uparrow}(t)$ (two-polarons). 
This population transfer occurs due to the resonance at $\Delta_+=9$ since 
$\abs{\Delta-\Delta_+}= 1.4\sim \Omega_{R0}$ yielding detuned Rabi-oscillations with $\Omega_R\approx 1.72$.
Notice that in the time-intervals $0<t<T_1/3$ and $2T_1/3<t<T_1$ $P_{\uparrow \downarrow}(t)$ oscillates in-phase 
[out-of-phase] with $P_{\uparrow \uparrow}(t)$ [$P_{\downarrow \downarrow}(t)$] and vice versa within $T_1/3<t<2T_1/3$. 
Furthermore, as in the previous case, the oscillation amplitude of $P_{\uparrow \downarrow}(t)$ is not fixed in 
the time-evolution which suggests once more the existence of weakly attractive induced impurity-impurity interactions 
mediated by the BEC medium. 

Next, we focus on strong impurity-medium interactions i.e. $g_{BI}=1.5$ where the TOC occurs and study 
$P_{aa'}(t)$ [Fig.~\ref{fig:two_body_prob} (c)] for an rf pulse possessing $\Omega_{R0}=1$ and e.g. $\Delta=-27.2$ 
namely at the region of the spectral resonance identified in $f(\Delta;\tau)$ [Fig.~\ref{fig:spectrum_weak_int} (c)]. 
We observe that for short evolution times ($t<10$), i.e. before the TOC occurs, the two bosonic impurities transfer 
from their spin $\ket{\downarrow\downarrow}$ configuration to a superposition of $\ket{\downarrow\uparrow}$ states. 
Later on, $10<t<30$, the impurities are majorly transferred with a slower rate to the $\ket{\uparrow\uparrow}$ state 
whose occupation subsequently ($t>30$) saturates to a finite value when also $\ket{\downarrow\downarrow}$ is almost 
completely depopulated. 
Interestingly, the population of the $\ket{\uparrow\downarrow}$ configuration tends to approach a fixed value due 
to the population of the $\ket{0,0}$ state while the $\ket{1,0}$ one does not contribute. 
This behavior can be explained by a blockade-like mechanism where the transfer of an impurity to the spin-$\uparrow$ 
state heavily distorts~\cite{Mistakidis_induced_cor,Mistakidis_pump} the bosonic background which consequently hinders 
the transfer of a second 
impurity to this state until ($t>10$) the first one escapes from the medium.

Turning to stronger rf pulses, i.e. $\Omega_{R0}=10$, and again residing close to the corresponding resonance 
where $\Delta_+\approx -26$ [Fig.~\ref{fig:spectrum_weak_int} (d)] the underlying spin-transfer processes are more 
involved [Fig.~\ref{fig:two_body_prob} (d)]. 
It becomes apparent that for $t<4$ a precession spin dynamics occurs where $P_{\downarrow \downarrow}(t)$ and 
$P_{\uparrow \uparrow}(t)$ oscillate out-of-phase with a non-constant amplitude. 
The latter indicates that the two-polaron state [see also Fig.~\ref{fig:density_strong_int_weak_dr} (e)] whose 
occupation is given by $P_{\uparrow \uparrow}(t)$ possesses a different energy from the single-polaron 
configuration [see $P_{\uparrow \downarrow}(t)$], which in turn suggests the presence of induced impurity-impurity 
interactions. For later times a portion of the spin-$\uparrow$ impurities 
[see also Fig.~\ref{fig:density_strong_int_weak_dr} (e)] escapes from the BEC environment resulting in a positive 
shift of the time-averaged population of the polaron states containing one and two spin-$\uparrow$ atoms. 
In the same time-interval $P_{\downarrow \downarrow}(t)$ and $P_{\uparrow \downarrow}(t)$ oscillate in-phase and out-of-phase with $P_{\uparrow \uparrow}(t)$. 
This behavior demonstrates the transfer of the trapped (within the BEC) atoms among their spin states and 
in particular between the $\ket{1,0}$ and the $\ket{1,1}$ ones. 

To further expose the presence of impurity-impurity correlations we utilize the spin correlation measure 
\begin{equation}
C_{\uparrow \downarrow}(\Delta;t)=4P_{\uparrow \uparrow}(t)P_{\downarrow \downarrow}(t)-P_{\uparrow \downarrow}^2(t),\label{spin_corel}
\end{equation} 
which takes values in the interval $[ -1, 1 ]$. 
In particular, if $C_{\uparrow \downarrow}(t)=0$ evinces the absence of correlations between the spin states of the impurities while for $C_{\uparrow \downarrow}(\Delta;t)>0$ [$C_{\uparrow \downarrow}(\Delta;t)<0$] signifies the bunching [anti-bunching] tendency 
of the impurities in the $\ket{\uparrow}$ and $\ket{\downarrow}$ states or equivalently in the $S_z=\pm 1$ spin configurations. 
The time-evolution of $C_{\uparrow \downarrow}(\Delta;t)$ for $g_{BI}=-0.5$ and $g_{BI}=1.5$ with varying detuning is presented 
in Figs.~\ref{fig:two_body_prob} (e) and (f) respectively. 
Focusing on $g_{BI}=-0.5$ [Fig.~\ref{fig:two_body_prob} (e)] it becomes apparent that in the vicinity of the attractive polaron 
resonance ($7.4<\Delta<10.4$), see also Fig.~\ref{fig:spectrum_weak_int} (e), $C_{\uparrow \downarrow}(\Delta;t)$ features an almost 
periodic behavior exhibiting bunching and anti-bunching of the impurities in the spin-$\uparrow$ and 
$\downarrow$ states. 
This process suggests the presence of impurity-impurity induced interactions. 
Turning to the motionally excited states of the polaron located at $5.2<\Delta<6.4$ the correlator $C_{\uparrow \downarrow}(\Delta;t)$ shows 
a fluctuating character remaining predominantly positive in the course of time thus indicating the two-body nature of these resonances. 

For strongly repulsive impurity-medium couplings where the TOC phenomenon takes place the response of $C_{\uparrow \downarrow}(\Delta;t)$ 
[Fig.~\ref{fig:two_body_prob} (f)] is more involved compared to the attractive interaction case. 
Indeed, close to the resonance at $\Delta_+^*\approx 26$ an anti-bunching is observed at the initial stages of the 
dynamics ($t<5$), as depicted in the inset of Fig.~\ref{fig:two_body_prob} (f), which 
is another manifestation of the blockade-like mechanism outlined previously, see also Fig.~\ref{fig:two_body_prob} (c). 
As time-evolves $C_{\uparrow \downarrow}(\Delta;t)>0$ due to the escape of the spin-$\uparrow$ impurity from the BEC background which allows for 
the excitation of the second one initially residing in the spin-$\downarrow$ state as also captured by $P_{\uparrow \downarrow}(t)$ 
[Fig.~\ref{fig:two_body_prob} (c)]. 
For longer times, $t>20$, $C_{\uparrow \downarrow}(\Delta;t)\to 0$ as a result of the population trapping to the spin-$\uparrow$ state due to the TOC. 
Regarding the lower-lying resonances at $\Delta>-26$ this blockade-like phenomenon is also evident since $C_{\uparrow \downarrow}(t)<0$ at
the initial time-interval of the impurities spin dynamics, see e.g. $13<t<30$ for $\Delta_+=-22.4$. 
Afterwards the impurities are bunched in the $\ket{\uparrow}$ state due to the TOC. 
In contrast, for $\Delta<-30$ the impurities experience a bunching tendency throughout the evolution.    
\begin{figure*}[ht]
\includegraphics[width=0.9\textwidth]{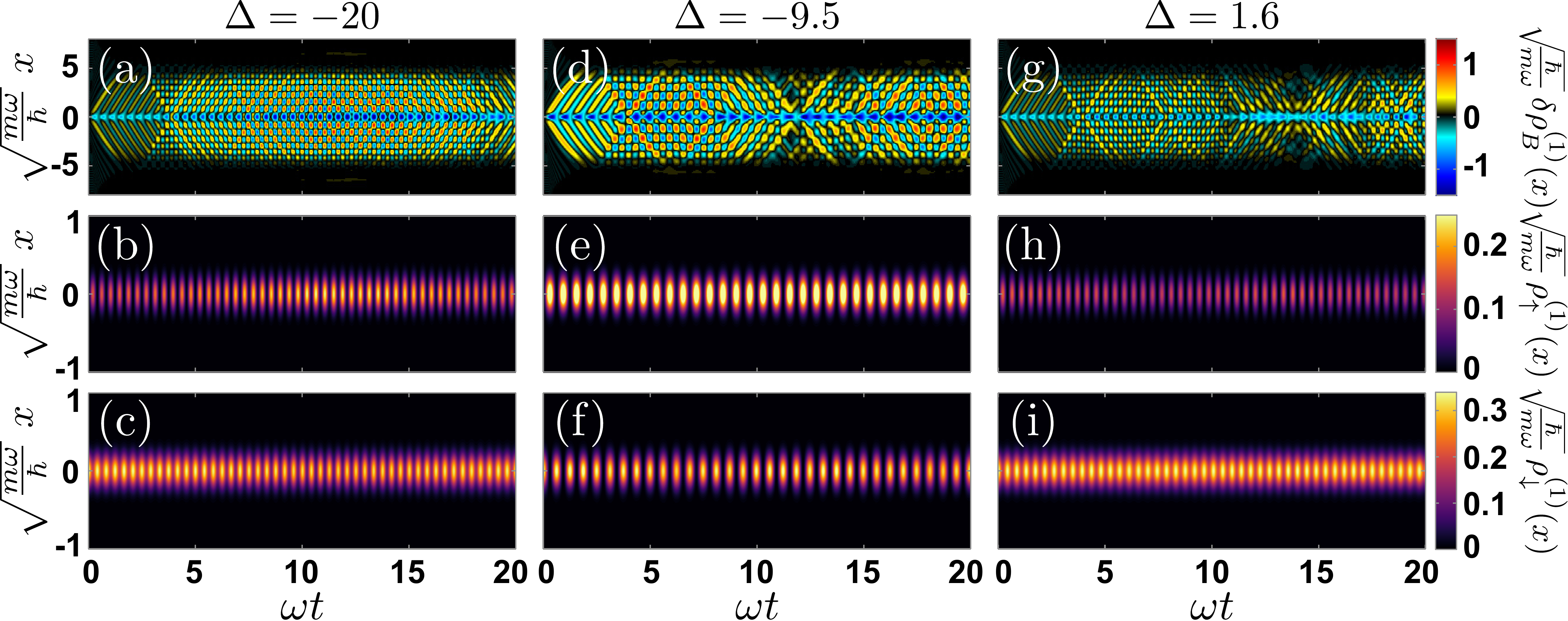}
\caption{(a), (d), (g) Temporal fluctuations of the one-body density of the BEC background 
$\delta \rho^{(1)}_B(x;t)=\rho^{(1)}_B(x;t)-\rho^{(1)}_B(x;0)$ 
for different detunings (see legends) and fixed bare Rabi frequency $\Omega_{R0}=10$. 
Density evolution (b), (e), (h) $\rho^{(1)}_{\uparrow}(x;t)$ and (c), (f), (i) $\rho^{(1)}_{\downarrow}(x;t)$ 
of a heavy single impurity 
for varying $\Delta$ (see legends) and constant $\Omega_{R0}=10$ undergoing Rabi-oscillations. 
Prominent interference phenomena between the impurity motion and the phononic excitations of the bath are clearly imprinted in $\delta \rho^{(1)}_B(x;t)$. 
In all cases, $m_I=(133/87)m_B$, $g_{B \downarrow}=0$, $g_{B\uparrow}\equiv g_{BI}=0.5$ are 
considered and the remaining parameters are 
the same as in Fig.~\ref{fig:spectrum_heavy_imp}. 
In all cases the pulse time corresponds to $\tau=T=100>t$ and the Thomas-Fermi radius of the bosonic bath is 
$R_{TF}\approx 4.2$.}
\label{fig:density_heavy} 
\end{figure*} 

\section{Heavy impurity: Bose polaron and associated interference processes}\label{sec:heavy_imp} 

Having analyzed in detail the spectral response of an impurity immersed in a BEC environment in the equal mass case, 
subsequently we intend to unravel the rf spectrum of a heavy impurity and contrast it with the results obtained for the mass-balanced case. 
To this end, we consider the experimentally relevant mixture of a single $^{174}$Yb impurity e.g. residing in the hyperfine states $\ket{F=2,m_F=2}$ and $\ket{F=2,m_F=1}$ and being immersed in a $^{7}$Li bosonic bath at $\ket{F=1,m_F=0}$ 
while the components are trapped in the same harmonic trap~\cite{Yb_exp}. 
For convenience we initially focus on repulsive impurity-bath interactions particularly $g_{BI}=0.5$ and 
inspect the rf signal for different bare Rabi frequencies, see Figs.~\ref{fig:spectrum_heavy_imp} (a)-(f). 

Regarding $\Omega_{R0}=1 \ll c/ \xi$ we observe a central polaronic peak at $\Delta_+^*\approx-8.8$ in $f(\Delta;\tau)$ showing a dominant 
Rabi frequency dictated by the relation 
$\Omega_R(\Delta)=\sqrt{(Z\Omega_{R0})^2+(\Delta-\Delta_+^*)^2}$ with $Z\approx 0.98$ as displayed 
in the inset of Fig.~\ref{fig:spectrum_heavy_imp} (a). 
Interestingly, a series of additional resonances appear in the spectrum $\bar{f}(\Delta)$ manifested as peaks for 
smaller detunings e.g. at $\Delta_+\approx -10$, $\Delta_+\approx -12$, $\Delta_+\approx -13.5$ and shallow 
depletions for a larger $\Delta$ located for instance at $\Delta_+\approx -6.2$, $\Delta_+ \approx -5.8$ and 
$\Delta_+\approx -4.2$. 
Their Rabi frequency [see the dashed box in the inset of 
Fig.~\ref{fig:spectrum_heavy_imp} (a)] is appreciably smaller compared to the one for $\Delta_+^*\approx -8.8$ since these 
motionally excited polaron states have a suppressed overlap with the initial non-interacting state or equivalently decreased residue. 
As we shall argue below these resonant peaks and depressions refer to the coupling of the impurity motion with the collective 
excitations, here the breathing mode, of its background. 
Of course, the effective potential approximation is not able to identify such mode-couplings since by construction 
it assumes that no medium excitations are induced. 
For this reason $\bar{f}(\Delta)$ within $V_{eff}(x;g_{BI})$ has a Lorentzian-like form with a single polaron 
resonance at $\Delta_+^*\approx -8.8$. 

The above-described composition of the spectral response remains robust and its features are amplified 
for a larger bare Rabi frequency, e.g. $\omega \ll \Omega_{R0}=10< c/ \xi$ illustrated in 
Figs.~\ref{fig:spectrum_heavy_imp} (c), (d). 
Indeed, an overall similar shape of both $f(\Delta;\tau)$ and $\bar{f}(\Delta)$ to the $\Omega_{R0}=1$ scenario is evident. 
Namely, there is a polaronic peak located at $\Delta_+^*\approx -10.3$ and side resonances manifested as peaks for 
$\Delta<\Delta_+$ [e.g. at $\Delta_+ \approx -17.1$, $\Delta_+ \approx -20 $, $\Delta_+ \approx -22.3 $] and depressions of 
$\bar{f}(\Delta)$ for $\Delta>\Delta_+$ [e.g. at $\Delta_+ \approx -0.4$, $\Delta_+ \approx 1.6$, $\Delta_+ \approx 4$]. 
Notice here that the frequency participating in the underlying spin-flip dynamics at $\Delta_+^*$ is 
$\Omega_R(\Delta)=\sqrt{(Z\Omega_{R0})^2+(\Delta-\Delta_+^*)^2}$, with $Z\approx 0.98$, see the inset of Fig.~\ref{fig:spectrum_heavy_imp} (c). 
As explained previously $V_{eff}(x;g_{BI})$ ignores the excitations of the bath and thus can not predict the side 
peaks of $\bar{f}(\Delta)$ present in the MB case. 
These side peaks and depressions of the spectrum arise also for a range of impurity-medium interaction strengths 
while their number is larger for more repulsive couplings as shown in Fig.~\ref{fig:spectrum_heavy_imp} (g). 
Also, we can readily infer that the position of the main polaronic resonance depends crucially on $g_{BI}$. 

The origin of this structure of the rf spectrum can be understood as follows. 
For fixed bare Rabi frequency $\Omega_{R0}$ the frequency $\Omega_R$ referring to the transfer of the impurity 
from its spin-$\downarrow$ to the spin-$\uparrow$ state increases as the detuning is shifted away from the polaronic 
resonance since $\Omega_R(\Delta)=\sqrt{(Z\Omega_{R0})^2+(\Delta-\Delta_+^*)^2}$ where $Z\approx 0.98$. 
As a consequence of this periodic transfer, the background density experiences a temporally varying force due to the 
finite $g_{BI}$ with an ``effective'' frequency $\sim \Omega_R$. 
Due to this perturbative force phononic excitations in the form of sound waves are induced in the 
medium~\cite{Marchukov_sw} for 
particular values of $\Omega_{R}$ and therefore for specific $\Delta$. 
Accordingly, these phononic excitations affect the impurity leading to its modified dressing with respect to the 
$\Delta=\Delta_+^*$ case signifying enhanced or suppressed polaron formation. 
This modification in the dressing of the impurity results in a modulation of the rf signal being manifested in the 
spectrum as an almost equidistant series of side peaks for $\Delta<\Delta_+^*$ and local depressions 
for $\Delta>\Delta_+^*$. 

To corroborate our above argumentation we showcase the density evolution of the spin states and the bosonic bath 
for specific detunings, corresponding to a peak ($\Delta_+\approx -20$), the main resonance ($\Delta_+^* \approx -9.5$) and a 
depression ($\Delta_+ \approx 1.6$) of the $\bar{f}(\Delta)$ with $\Omega_{R0}=10$ and $g_{BI}=0.5$. 
More precisely, for the BEC medium we monitor its density fluctuations i.e. 
$\delta \rho_B^{(1)} (x;t)= \rho_B^{(1)}(x;t)-\rho_B^{(1)}(x;0)$, see also Eq.~(\ref{eq:single_par_den_matr}), in 
order to subtract its initial (ground state) background with $\rho^{(1)}(0,0)\approx 16$ and render the emission of 
the phononic excitations of amplitude $\sim$1 visible. 
Monitoring $\rho_{\uparrow}^{(1)}(x;t)$ and $\rho_{\downarrow}^{(1)}(x;t)$ we observe that independently of the 
detuning a Rabi-oscillation, i.e. a periodic population transfer from the $\ket{\uparrow}$ to the $\ket{\downarrow}$ 
state, takes place [see Figs.~\ref{fig:density_heavy} (b), (c), (e), (f), (h), (i)] with a frequency according 
to $\Omega_R(\Delta)=\sqrt{(Z\Omega_{R0})^2+(\Delta-\Delta_+^*)^2}$ where $Z\approx 0.98$. 
This process signifies the periodic polaron generation in the system. 
Most importantly the bosonic bath experiences a more involved dynamical response which is crucially affected 
by the value of the detuning [Figs.~\ref{fig:density_heavy} (a), (d), (g)]. 
The spin-$\uparrow$ impurity motion imprints a prominent sound wave emission for $t<4$ as it can be seen 
in $\delta \rho_B^{(1)}(x;t)$~\cite{Mukherjee_pulse,Marchukov_sw} since here $\Omega_{R}(\Delta) \sim \Omega_{R0} \sim c/ \xi$ 
with the latter referring to the characteristic phonon energy scale. 
Therefore, the corresponding side resonance structure of $\bar{f}(\Delta)$ [Fig.~\ref{fig:spectrum_heavy_imp} (d)] 
is accordingly attributed to the locally adiabatic character of the spin-transfer. 
In particular, these sound waves emanating from the trap center travel towards its edges and are reflected back. 
Subsequently, a detuning dependent interference pattern occurs within the spatial extent of the bosonic bath. 
Indeed, for detunings corresponding to a side peak (depletion) of $\bar{f}(\Delta)$ the sound waves interfere 
constructively around $x=0$ yielding pronounced oscillations of the bath density around the trap center, see 
Figs.~\ref{fig:density_heavy} (a) and (g) respectively. 
These density oscillations in turn periodically modulate the resonant frequency since 
$\Delta_+^*\propto g_{BI} \rho^{(1)}_B(0;t)$, resulting in an enhanced (reduced) polaron formation 
for $\Delta>\Delta_+^*$ ($\Delta<\Delta_+^*$), see also 
remark in Ref.~\cite{comment2}. 

Further increasing the bare Rabi frequency of the rf field such that $\Omega_{R0}\gg c/ \xi$ and operating with 
$g_{BI}=0.5$ practically leads to the diabatic spin-transfer regime. 
This fact is explicitly demonstrated in $f(\Delta;\tau)$ and $\bar{f}(\Delta)$ upon applying a pulse 
of $\Omega_{R0}=40$ [Fig.~\ref{fig:spectrum_heavy_imp} (c)]. 
It becomes evident that $f(\Delta;\tau)$ features a polaronic resonance at $\Delta_+^*\approx -15.2$ and Rabi oscillates 
with a frequency of $\Omega_R=\sqrt{(Z\Omega_{R0})^2+(\Delta-\Delta_+^*)^2}$ and $Z\approx 0.98$, see the inset of 
Fig.~\ref{fig:spectrum_heavy_imp} (c). 
Since in this case the mode-coupling resonances (i.e. the side-peaks) are smeared out, the effective potential 
approximation can sufficiently predict the resultant rf spectrum $\bar{f}(\Delta)$ [Fig.~\ref{fig:spectrum_heavy_imp} (f)] because impurity-BEC correlations hardly develop within the small timescale that the impurity resides in the spin-$\uparrow$ state. 
Similar conclusions, in this strongly driven regime, can also be drawn for the spectral configuration when 
considering different impurity-medium couplings, see Fig.~\ref{fig:spectrum_heavy_imp} (h). 
For instance, the form of $\bar{f}(\Delta)$ remains the same when $g_{BI}=\pm 0.5$ with the resonance being shifted 
to positive detunings if $g_{BI}=-0.5$. 
Interestingly, for large impurity-bath repulsions e.g. $g_{BI}=1.5$ additional side peaks and depressions build upon 
the spectral line $\bar{f}(\Delta)$, a phenomenon that is reminiscent of the previously discussed weak pulse scenario. 
Especially, the peaks originate from the presence of strong interactions leading to the build-up of enhanced impurity-bath correlations.

\section{Summary and Outlook}\label{sec:conclusions}

We have emulated reverse rf spectroscopy of one-dimensional and harmonically trapped Bose polarons at  
different spin-transfer regimes ranging from the adiabatic to the diabatic limit as quantified by the intensity of the applied rf field. 
In this sense, we expose the alterations 
of the impurity spectral response stemming from fully adiabatic and diabatic rf pulses providing also insights into the 
intriguing intermediate spin-transfer regime termed as locally adiabatic. 
This investigation enables us to discern the emergent correlation induced mechanisms of the underlying spin-flip 
dynamics from single-particle ones and to analyze the origin of the formation of polaronic excitations or 
mode-couplings of the latter with collective excitations of the background. 
The location of the identified polaronic resonances depends on the impurity-bath coupling for fixed bare Rabi 
frequency of the rf field and vice versa. 
The analysis is based on a variational non-perturbative treatment of the impurities nonequilibrium dynamics that allows us to capture 
the interparticle correlations of the multicomponent setup. 
Comparisons with an effective potential approach and the diabatic approximation are performed yielding further insights into the nature of the observed spectral resonances.

Focusing on weakly attractive or repulsive impurity-medium interactions we exemplify the formation of coherent attractive 
or repulsive Bose polarons respectively. 
In particular, for pulses characterized by a bare Rabi frequency smaller or equal to the trap one we identify a multitude of polaronic resonances in the spectrum referring to lower and motionally excited states of the generated quasi-particle as well as couplings of the impurity motion with the excitations (breathing mode) of the bosonic bath. 
These excited states are associated, for instance, with a multihump shape of the impurity density and the emission of 
sound waves in the bosonic environment. 
In the case of repulsive interactions the involved spin-mixing dynamics contains a variety of frequencies while for 
attractive couplings it only involves a single one. 
An effective potential picture is also constructed to interpret the nature of the polaron excitation processes which are 
inherently related to a decreasing quasi-particle residue for increasing polaron energy. 
Considering two bosonic impurities the existence of induced interactions for attracting impurity-medium couplings 
can be inferred from the small shift of specific spectral resonances. 
Increasing the bare Rabi frequency such that it exceeds the one set by the energy scale of phononic excitations, the transition to the diabatic spin-transfer regime is verified with the spectrum exhibiting a single polaronic resonance and its shape being adequately predicted by 
the diabatic approximation. 

Turning to strongly repulsive impurity-bath coupling strengths and for weak bare Rabi frequencies we probe the 
dynamical decay of the Bose polaron, termed TOC and manifested by the prominent saturation 
trend of the rf signal for long pulse times. 
The resultant spectral response features a variety of resonances evincing the existence of distinct superpositions 
of energetically lower-lying polaronic excitations. 
The latter can be intuitively understood in terms of an effective potential approximation which accurately predicts 
the locations of the corresponding spectral peaks. 
Inspecting the spatiotemporal evolution of the spin densities we observe the build up of their multihump shape supporting 
the excitation process of the impurity. 
Remarkably, the decay of the Bose polaron is associated with an energy transfer from the impurity to the 
BEC background and a suppressed coherence of the former, with both processes being enhanced for a weaker pulses. 
In the case of two impurities slight shifts in the position of specific spectral peaks with respect to the 
ones of a single impurity are indicative of the presence of induced impurity-impurity interactions. 
To further demonstrate the importance of induced interactions we unravel a blockade-like phenomenon by studying 
the populations of one and two polaron states. 
Utilizing a larger bare Rabi frequency of the rf field which overcomes the characteristic one of phononic excitations 
we can systematically reach the 
diabatic spin-transfer regime represented by a Lorentzian distribution and featuring a single polaronic resonance. 
Interestingly, the range of Rabi frequencies which can yield a multi-peaked spectral response is broader here when 
compared to weaker interactions owing to the emergence of the locally adiabatic regime.

The spectrum of an impurity heavier than the atoms of its bath is drastically modified when compared to the equal mass scenario. 
Here, the spectral response possesses a complex structure independently of the impurity-medium coupling and 
the bare Rabi frequency of the rf field such that we do not enter the diabatic regime. 
It consists of a polaronic resonance accompanied by a series of equidistant side peaks and depletions. 
These side resonances signify the coupling of the impurity spin dynamics with the sound wave excitations of 
the background yielding pronounced interference phenomena which are especially prominent in the locally adiabatic spin-transfer regime. 
Naturally, they can not be captured with an effective potential approximation, which does not account for the 
excitations of the bath, while their number is larger for 
increasingly repulsive interactions. 
For sufficiently large Rabi frequencies we reach the diabatic spin-transfer regime where the side peaks and depressions are 
smeared out and the spectral response has a Lorentzian shape showing a single polaronic resonance in perfect agreement 
with the diabatic approximation. 

There is a plethora of interesting extensions of the present results that are worth being pursued in the future. 
Certainly, the generalization of the current findings to higher spatial dimensions and to the case of both Bose 
and Fermi impurities exposing this way also the role of the different statistics is extremely desirable. 
Moreover, it is important to study the robustness of the emergent spin-flip dynamics in the current and 
higher-dimensional settings in the presence of temperature effects~\cite{Dzsotjan,Tajima,Liu_rf1,Tajima_thermal}. 
Another intriguing direction would be to unravel the rf spectrum at different spin-transfer regimes 
of magnetic quasi-particles such as magnetic polarons or Bose polarons embedded in a spinor bosonic medium. 
Here, also the investigation of the properties of the rf dressed quasi-particles, e.g. their lifetime, residue, 
effective mass and induced interactions would be a relevant topic.

\appendix

\section{Diabatic approximation}\label{app:diabatic_approx}

As discussed in the main text, Sec.~\ref{sec:hamiltonian_protocol}, for rf pulses characterized by a bare Rabi 
frequency $\Omega_{R0}\gg c/ \xi$ we enter the {\it diabatic spin-transfer regime} and impurity-medium correlations 
are expected to be suppressed since the spatial part of the MB wave function can not adapt to the small timescale of the pulse. 
Indeed, in this case where $\Omega_{R0} \gg c/ \xi$ it is possible to invoke the so-called diabatic approximation 
for approximating the injection spectrum of the impurities being embedded in the BEC environment. 
The key assumption of this approximation is that the kinetic energy of both the impurities and the medium is 
negligible within the duration of the pulse and can therefore be safely dropped. 
Note that the diabatic approximation does not rely on the omission of the kinetic energy of the system's initial state provided that the corresponding momentum satisfies $\braket{p_{\sigma}(0)} \ll m \Omega_{R0} \xi / \pi$ with $\sigma=B,I$. 
In fact, for $\Omega_{R0} \gg c/ \xi$ the resulting Rabi oscillation of the impurities between their spin-$\downarrow$ 
and spin-$\uparrow$ states occurs at a much faster timescale i.e. $\sim1/\Omega_{R0}$ when compared to the relative movement between the bath and the impurity atoms being of the order of $\sim \xi/c$. 
The latter corresponds to the characteristic timescale where the elementary excitations of the BEC can appear. 
Accordingly, it is legitimate to treat the atoms as stationary during the injection pulse. 
The diabatic approximation greatly simplifies the MB Hamiltonian of Eq.~(\ref{Htot_system}) whose approximate form reads
\begin{equation}
\begin{split}
\hat{H}_{\rm diab}&=\int {\rm d}x~ \bigg\{\frac{1}{2} m \omega^2 x^2 \bigg[ 
\hat{\Psi}^{\dagger}_B(x) \hat{\Psi}_B(x) +
\hat{\Psi}^{\dagger}_\uparrow(x) \hat{\Psi}_\uparrow(x) \\&+
\hat{\Psi}^{\dagger}_\downarrow(x) \hat{\Psi}_\downarrow(x) \bigg]
+ \frac{g_{BB}}{2} \hat{\Psi}^{\dagger}_B(x) \hat{\Psi}^{\dagger}_B(x) \\& \times \hat{\Psi}_B(x) \hat{\Psi}_B(x)
+ g_{BI} \hat{\Psi}^{\dagger}_B(x) \hat{\Psi}^{\dagger}_\uparrow(x) \\& \times \hat{\Psi}_\uparrow(x)
\hat{\Psi}_B(x)\bigg\}
+ \frac{\hbar \Delta}{2} \hat{S}_x 
+ \frac{\hbar \Omega_{R0}}{2} \hat{S}_y. 
\end{split}
\label{eq:hamilt_diab}
\end{equation}
Here, the parameters $\omega$, $g_{BB}$, $g_{BI}$, $\Delta$ and $\Omega_{R0}$ denote the trapping frequency, individual 
interparticle interactions, detuning and bare Rabi frequency respectively and are essentially the same as the ones appearing in Eq.~(\ref{Htot_system}). 
Importantly, notice that since the interparticle interaction is zero-range, $\hat{H}_{\rm diab}$ is diagonal in real space. 
A crucial corollary of the above fact is that within $\hat{H}_{\rm diab}$ the state of the bath cannot change during the time-evolution. 
Turning to the impurities, their spatial distribution as captured by their total density, 
$\rho^{(1)}_{I}(x)=\rho_{\uparrow}^{(1)}(x)+\rho_{\downarrow}^{(1)}(x)$, remains constant in the course of the dynamics. 
Nevertheless, the spatial distribution of their spin-$\uparrow$ and spin-$\downarrow$ components is allowed to evolve 
due to the finite $\Omega_{R0}$. 
To further simplify our analysis we assume here that similarly to the situation described in the main text the bath and 
the impurity are initially non-interacting, since $\langle \Psi (0) | \hat{S}_z | \Psi (0) \rangle = -1$. 
Hence, the impurity-medium entanglement is initially zero. 
This implies that the state of the bath and the impurities possesses the product form $| \Psi(0) \rangle = |
\Psi_B (0) \rangle \otimes |\Psi_I(0) \rangle$. 
As a consequence $\hat{H}_{\rm diab}$ can be projected to the subspace where the bath remains in its initial state, yielding 
the following Hamiltonian for the impurity degree-of-freedom
\begin{equation}
\begin{split}
\hat{P}_{B} \hat{H}_{\rm diab} \hat{P}_{B}&=E_0 + \frac{\hbar \Omega_{R0}}{2} \hat{S}_x\\&+\int {\rm d}x~
 \frac{\hbar}{2} \left(  \Delta + \frac{g_{BI} \rho_B^{(1)}(x)}{2} \right) \hat{S}_z (x),
\end{split}
\label{eq:hamilt_proj}
\end{equation}
where $\hat{P}_{B}=| \Psi_B (0) \rangle \langle \Psi_B (0) |$ is the corresponding projection operator into the
bath initial state and $\rho_B^{(1)}(x)$ refers to the one-body density of the bosonic environment. 
The term $E_0$ refers to a constant energy shift which, of course, does not contribute to the dynamics and reads
\begin{equation}
\begin{split}
E_0=&\frac{1}{2} m \omega^2 \int {\rm d}x~x^2 \left( \rho_B^{(1)}(x) + \rho_I^{(1)}(x) \right)
\\&+\frac{g_{BB}}{2} \int {\rm d}x~\rho_B^{(2)}(x,x). 
\end{split}
\end{equation}
In this expression, $\rho_I^{(1)}(x)=\rho_{\uparrow}^{(1)}(x)+\rho_{\downarrow}^{(1)}(x)$ is the impurities one-body 
density and $\rho_B^{(2)}(x,y)$ designates the two-body density of the bath atoms. 
The projected Hamiltonian, Eq.~(\ref{eq:hamilt_proj}), can be interpreted as an ensemble of continuously many
spin-$1/2$ particles that are non-interacting among themselves. 
This ensemble is subjected to a homogeneous Rabi coupling of amplitude $\Omega_{R0}$ that possesses a spatially dependent 
detuning, namely $\Delta(x)=\Delta + g_{BI} \rho^{(1)}_B(x)/2$. 
Therefore, the time-evolution of the system according to the approximate Hamiltonian Eq.~(\ref{eq:hamilt_diab}) can be 
expressed in terms of the solution 
of a Rabi-coupled two level system~\cite{Scully}. 
Indeed, the probability to find a spin-$\uparrow$ impurity at position $x$ after a rectangular pulse of
duration $t$ is 
\begin{equation}
\begin{split}
P_{\uparrow}&(x,t;\Delta)=\frac{1}{N_I} \frac{\Omega_{R0}^2 \rho^{(1)}_I(x)}{\Omega_{R0}^2+\left(\Delta
+ \frac{g_{BI} \rho_B^{(1)}(x)}{2} \right)^2} \\& \times \sin^2 \left[\sqrt{\Omega_{R0}^2+\left(  \Delta
+ \frac{g_{BI} \rho_B^{(1)}(x)}{2} \right)^2} \frac{t}{2} \right].
\end{split}
\end{equation}
Furthermore, the rf signal can be evaluated by the integration of the above-mentioned probability in space, 
i.e. $f(\Delta,\tau)=\int {\rm d}x~P_{\uparrow}(x,t;\Delta)$ and depends only on the one-body densities of the 
bath ($\rho_B^{(1)}(x)$) and impurity ($\rho^{(1)}_I(x)$) atoms.

Particularly, for the system prepared in its ground state with $g_{BB}>0$ and $\langle \Psi (0) | \hat{S}_z | \Psi (0) \rangle = -1$ 
as examined herein, the BEC density is well-described by the Thomas-Fermi profile $\rho_B^{(1)}(x)=\frac{m \omega^2}{2 g_{BB}}(R_{TF}^2-x^2)$, 
where $R_{TF} = (\frac{3g_{BB} N_B}{2 m \omega^{2}})^{\frac{1}{3}}$ refers to the Thomas-Fermi radius. 
On the other hand, the impurity density possesses a Gaussian form i.e. 
$\rho_I^{(1)}(x)=\frac{N_I }{\sqrt{\pi} \alpha} e^{-\frac{x^2}{\alpha^2}}$, with $\alpha=\sqrt{\hbar/(m \omega)}$ being the 
length of the external confinement. 
In this case, the averaged rf signal is composed of two parts namely, $\bar{f}(\Delta)=\bar{f}_1(\Delta)+\bar{f}_2(\Delta)$. 
The first term corresponds to the contribution of the impurity density within the spatial region occupied by the BEC and 
refers to the integral
\begin{equation}
\bar{f}_1(\Delta)=\frac{1}{2 \sqrt{\pi}} \int_{-R_{TF}/\alpha}^{R_{TF}/\alpha} {\rm d}y~e^{-y^2}
\frac{\tilde{\Omega}^2}{\tilde{\Omega}^2+(y^2-\tilde{\Delta})^2}. 
\label{eq:first_fd}
\end{equation}
Here, $\tilde{\Omega}(\Omega_{R0})=\frac{2 g_{BB} \Omega_{R0}}{g_{BI} \omega}$ and 
$\tilde{\Delta}(\Delta)=\frac{2g_{BB}}{g_{BI}} [ \frac{\Delta}{\omega} + \frac{1}{2} \frac{g_{BI}}{g_{BB}} (\frac{R_{TF}}{\alpha})^2]$. 
In addition, the second term corresponds to a Lorentzian shaped peak at $\Delta=0$, namely
\begin{equation}
\bar{f}_2(\Delta)=\frac{1}{2} \left[1 - {\rm erf} \left(\frac{R_{TF}}{\alpha}\right) \right]
\frac{\Omega_{R0}^2}{\Omega_{R0}^2+\Delta^2},
\label{eq:second_fd}
\end{equation}
with ${\rm erf}(x)$ being the error function. 
This term essentially accounts for the impurities lying beyond the Thomas-Fermi radius of their medium. 
Typically the BEC is rather extended, i.e. $R_{TF}>\alpha$, resulting to the contribution of $\bar{f}_2(\Delta)$ being rather small. 
More specifically, for $N_B=100$, $g_{BB}=0.5$ employed in Sec.~\ref{sec:hamiltonian_protocol}, $R_{TF} \approx 4.2$ and accordingly 
the amplitude of this Lorentzian peak is negligible $\bar{f}_2(\Delta) < 1.4 \times 10^{-9}$. 
Regarding the portion of the impurities lying within the BEC, notice that in the regime of the diabatic 
approximation ($\Omega_R \gg c/ \xi$) $\tilde{\Omega} \gg 1$ holds allowing for the calculation of the 
integral Eq.~(\ref{eq:first_fd}) which by employing a Taylor expansion yields 
\begin{equation}
\begin{split}
\bar{f}_1(\Delta)&=\frac{1}{2} \frac{\tilde{\Omega}^2}{\tilde{\Omega}^2+\tilde{\Delta}^2} \bigg[
1 +\frac{\tilde{\Delta}}{\tilde{\Omega}^2+\tilde{\Delta}^2} + \frac{3}{8} 
\frac{3 \tilde{\Delta}^2 - \tilde{\Omega}^2}{( \tilde{\Omega}^2 +\tilde{\Delta}^2 )^2} 
\\& +\mathcal{O} \left(\frac{\tilde{\Delta}^3}{( \tilde{\Omega}^2 +\tilde{\Delta}^2)^3},
\frac{\tilde{\Omega}^3}{( \tilde{\Omega}^2 +\tilde{\Delta}^2)^3} \right) 
\bigg].
\label{eq:Taylor_first_fd}
\end{split}
\end{equation}
Eq.~(\ref{eq:Taylor_first_fd}) dictates that within the diabatic regime an almost perfectly Lorentzian peak is
expected to be centered around $\tilde{\Delta}_0 \approx -0.5$, or equivalently 
$\Delta_0 \approx -\frac{\omega}{2} \frac{g_{BI}}{g_{BB}}\left[\left(\frac{R_{TF}}{\alpha} \right)^2 - \frac{1}{2} \right]$, with 
half width at half maximum given by $\gamma \approx \Omega_{R0}$. 
Notice the linear dependence of the width of the peak $\gamma$ on the applied Rabi frequency $\Omega_{R0}$ stemming from the 
significant power broadening exhibited in the rf spectrum within this regime.

\section*{Acknowledgements} 
%S. I. M. gratefully acknowledges financial support in the framework of the Lenz-Ising Award of the 
%University of Hamburg. 
%This work is supported by the Cluster of Excellence “CUI: Advanced Imaging of Matter” of 
%the Deutsche Forschungsgemeinschaft (DFG)-EXC 2056-project ID 390715994. 
S. I. M. gratefully acknowledges financial support in the framework of the Lenz-Ising Award of the University of Hamburg. 
This work is supported by the Cluster of Excellence 'Advanced Imaging of Matter' of the Deutsche Forschungsgemeinschaft (DFG) - EXC 2056 - project ID 390715994. 
This work has been funded by the Deutsche Forschungsgemeinschaft (DFG, German Research Foundation) – SFB-925 – project
170620586.
%P.S. gratefully acknowledges financial support by the Deutsche Forschungsgemeinschaft 
%(DFG) in the framework of the SFB 925 ``Light induced dynamics and control of correlated quantum
%systems''. 
F. G. acknowledges funding by the Deutsche Forschungsgemeinschaft (DFG, German Research Foundation) under 
Germany’s Excellence Strategy -- EXC-2111 -- 390814868. 
H.R.S. acknowledges the support from the NSF through a grant for ITAMP at Harvard University.

{}


\begin{thebibliography}{60}
\bibitem{comment1} Note that the use of $\rho^{(1)}_B(x)$ is justified in our case since the shape of the bosonic gas is not strongly 
distorted during the 
dynamics and it adequately corresponds to a Thomas-Fermi profile.
\bibitem{comment2} A similar spectral structure to the one observed in Figs.~\ref{fig:spectrum_heavy_imp} (d), (f) can be unveiled by the simple two-level model $\hat{H}=-(\hbar/2)[\Delta_+^* -\delta \sin(\omega t)]\hat{S}_z+(\hbar\Omega_{R0}/2) \hat{S}_x$, with $\omega >\Omega_{R0}$ and $\delta \ll \Omega_{R0}$.
\end{thebibliography}
\end{document}